\documentclass[a4paper,11pt]{article}
\pdfoutput=1 

\usepackage{jcappub} 

\usepackage[T1]{fontenc} 

\title{\boldmath Primordial Black Hole Formation in a Dust Bouncing Model}

\usepackage[english]{babel}
\usepackage{braket}
\usepackage{amsmath}
\usepackage{amsfonts}
\usepackage{amssymb}
\usepackage{physics,slashed}
\usepackage{graphicx}
\usepackage[toc,page]{appendix}
\usepackage{hyperref}
\usepackage[noabbrev]{cleveref}

\usepackage{xcolor}
\usepackage{mdframed}

\usepackage[hmarginratio=1:1]{geometry}

\renewcommand{\vec}{\mathbf}

\newcommand{\dpar}[1]{\left(#1 \right)} 
\newcommand{\dcol}[1]{\left[#1 \right]} 
\newcommand{\dcha}[1]{\left\{#1 \right\}} 

\author[a,1]{E.J. Barroso,}\note{Corresponding author.}
\author[a]{L.F. Demétrio,}
\author[a]{S.D.P. Vitenti,}
\author[b]{Xuan Ye}

\affiliation[a]{Physics Department, Universidade Estadual de Londrina, \\Campus Universitário, CEP 86057-970, Londrina, Brasil,}
\affiliation[b]{ Department of Astronomy, Key Laboratory for Researches in Galaxies and Cosmology,
   School of Astronomy and Space Sciences, University of Science and Technology of China,
   96 JinZhai Road, Hefei, Anhui, 230026, China}

\emailAdd{eduardo.jsbarroso@uel.br}
\emailAdd{demetrio.luizfelipe@uel.br}
\emailAdd{vitenti@uel.br}
\emailAdd{yyyyy@ustc.edu.cn}

\abstract{Linear scalar cosmological perturbations have increasing spectra in the
	contracting phase of bouncing models. We study the conditions for which these
	perturbations may collapse into primordial black holes and the hypothesis that these
	objects constitute a fraction of dark matter. We compute the critical density
	contrast that describes the collapse of matter perturbations in the flat-dust bounce
	model with a parametric solution, obtained from the Lemaitre-Tolman-Bondi metric
	that represents the spherical collapse. We discuss the inability of the Newtonian
	gauge to describe perturbations in contracting models as the perturbative hypothesis
	does not hold in such cases. We carry the calculations for a different Gauge choice
	and compute the perturbations' power spectra numerically. Finally, assuming a
	Gaussian distribution, we compute the primordial black hole abundance with the
	Press-Schechter formalism and compare it with observational constraints. From our
	analysis, we conclude that the primordial black hole formation in a dust-dominated
	contracting phase does not lead to a significant mass fraction of primordial black
	holes in dark matter today.}

\begin{document}
\maketitle
\flushbottom

\section{Introduction}
\label{sec:intro}

Primordial Black Holes (PBHs) are believed to have been formed in the early universe
through the collapse of density fluctuations~\cite{Zel1967, Hawking1971, Hawking1974,
	Carr1974}. Due to their formation mechanisms, in the inflationary scenario, PBHs can
have a wider range of masses from around $M \sim 5\times 10^{-29} M_\odot$ if formed at
Planck time $t = 10^{-43}s $ and $M \sim 10^5 M_\odot$ if formed at $t\sim 1s$,
resulting in several different effects. The study of primordial black holes has yielded
insights into various phenomena. They could have contributed to the cosmological density
parameter~\cite{Carr1975, Hee1996}. Their influence in the Cosmic Microwave Background
Radiation (CMB) was studied in Ref.~\cite{Ricotti2008} and some recent works explored
the connection between evaporating PBHs and gravitational waves~\cite{Dom2021}. There
also have been discussions about whether some measurements of gravitational waves could
be attributed to primordial black holes~\cite{Wang2022}. Undoubtedly, the possibility
that primordial black holes constitute a significant fraction of cold dark matter
remains a focal point in current research~\cite{1975Natur}, since they are labeled as
non-baryonic as their formation takes place before the Big Bang Nucleosynthesis
(BBN)~\cite{Cyburt2003}. Recent observations of merging binary black holes with
unexpected mass ranges by the LIGO/Virgo collaborations indicate that they could have
originated from PBHs~\cite{Abbott2016, Abbott2019}.

Various mechanisms can result in primordial black hole formation (see
Ref.~\cite{Escriva2023} for a review). They can originate from the collapse of large
isocurvature perturbations of cold dark matter~\cite{Passaglia2022}, first-order phase
transitions~\cite{Khlopov1998, Liu2022}, critical collapse of matter
perturbations~\cite{Hawking1971, Carr1974}, and others. Based on the study of critical
phenomena and simulations, the formation mass of black holes in this context will
heavily depend on the cosmological model and the shape of perturbations
~\cite{Niemeyer1998}, which will directly impact the critical threshold $\delta_c$.
Thus, the density contrast and its critical value must be carefully studied.

In the inflationary scenario~\cite{Starobinskii1979, Guth1981, Bardeen1983, Linde1982},
several studies have analyzed the formation of PBH at the end of
inflation~\cite{Bullock1997, Yokoyama1998, Josan2010, Ballesteros2018, Wang2024} and in
the reheating phase~\cite{Carr2018, Martin2020} either as a probe for the inflationary
model or to analyze the abundance of PBH in dark matter. When confronting the results
with combined probes from observational data, it was found that the only acceptable PBH
mass range that allows these objects to entirely constitute dark matter is
$10^{-16}M_\odot - 10^{-12}M_\odot$~\cite{Carr2022, Villanueva2021}. Nonetheless, other
mass ranges that lead to a smaller fraction of dark matter have intriguing impacts to be
considered. For instance, in the mass range $10^6M_\odot$ to $10^{10} M_\odot$ where
PBHs only represent $0.1\%$ of DM, they could play a role in generating supermassive
black holes. In \cite{Garcia2017} it was also concluded that the possibility of dark
matter being constituted by supermassive primordial black holes is viable. However,
these results led researchers to study PBH formation in other cosmological models to
check the dependency of the mass constraints on the chosen models.

As is well known, the inflationary paradigm alleviates the initial condition problems of
the $\Lambda$-CDM model but does not completely solve
them~\cite{Guth1981,nelson2021bouncing, PatrickReview2}. For instance, one still needs
to assume that a small patch of spacetime was of FLRW type and evolved to become the
universe that we now observe, which is modeled via perturbation
theory~\cite{covariant_bardeen}. Alternative models to inflation have been considered,
mainly bouncing models (see~\cite{Gasperini1993, Gasperini1994, Lyth, Finelli,
	Wands1999, Brandenberger2001, Peter2002, Hwang2002, Vitenti2012, Vitenti2013,
	PatrickReview1} for extensive reviews on bouncing cosmology), which focus on solving the
singularity problem by introducing a contracting phase connected to the present
expanding phase trough a minimal scale factor: a bounce. Classical bouncing models,
where exotic matter or modifications of GR are considered, have been analyzed
extensively in the literature, mainly in Refs.~\cite{PatrickReview1, PatrickReview2}. In
these works, one sees that classical bounces are non-trivial to implement and lead to
undesired features such as instabilities and new singularities, which are associated
with the violation of the null energy condition.

One may also consider quantum bounces, where the quantization of gravity itself
eliminates the primordial singularity. In particular, since quantum bounces do not make
use of exotic matter such as the inflaton, there is no need for a reheating phase. This
has been achieved in previous works in the framework of Canonical Quantum
Gravity~\cite{nelson_peter_bouncing_original}. Other relevant proposals are Loop Quantum
Cosmology \cite{loop_quantum_gravity_perturbations_application,loop_phenomenology} and
String Gas Cosmology \cite{PatrickReview2, PatrickReview1}.

Some recent studies have analyzed the formation of PBH in bouncing
models~\cite{Carr2011, Corman2022, Chen2017, Chen2023, Quintin2016, Banerjee2022,
	Papanikolaou2024}. In this context, it is believed that the long duration of the
contracting phase plus the diminishing scale factor may lead to an enhancement of PBH
formation, resulting in a more significant contribution to the DM density. In
\cite{Quintin2016} it is concluded that a dust-dominated bouncing universe ($w > 1/3$,
being $w$ the fluid's equation of state) is robust against the formation of such
primordial black holes. In contrast, for a matter-dominated universe ($w \ll 1$), their
formation becomes relevant. A similar conclusion is achieved in \cite{Chen2023}, where
they got an enhanced production of PBHs near the bouncing point and utilized the
abundance of PBHs in DM to constrain the bouncing model. However, in the same work, it
is stated that these constraints are still not well understood due to a lack of
precision in numerical computations.

Furthermore, it is still not quite clear how to properly define the critical threshold
for PBHs in bouncing cosmology. Different from the inflationary scenario, the
contracting phase dynamics lead to growing perturbations that will collapse before
becoming super-Hubble. To circumvent this problem, in~\cite{Quintin2016} it is used the
argument that the black holes must be larger or equal to the Schwarzschild radius for
the given formation mass seeded by the perturbations. However, there is still the need
for a more precise definition of the critical contrast in the bouncing scenario.
Furthermore, the gauge definition used in previous works may not be the best choice as
it leads to large increasing spectra and thus a miscalculation of the energy-density
perturbations~\cite{vitenti2012large}.

In this work, we study the formation of PBH in a quantum dust non-singular bounce and
the hypothesis that these structures constitute a fraction of DM today. We consider the
single barotropic fluid quantum bouncing model developed in
\cite{nelson_peter_bouncing_original, nelson2000bohm, nelson2021bouncing} using
Canonical Quantum Gravity, which is a conservative approach to the quantization of
General Relativity and should hold as an effective theory up to a certain energy scale
\cite{nelson2021bouncing}. We shall focus on the critical collapse of matter
perturbations characterized by the density contrast $\delta$, such that the
perturbations collapse to form black holes when they achieve a given threshold ($\delta
	> \delta_c$). To perform this analysis, we need to compute the spectra of the
perturbations and the critical threshold needed for the distribution of PBHs. We compute
the critical threshold in a more detailed approach, using the Tolman-Bondi-Lemaitre
metric~\cite{Tolman1934, Lemaitre1933} in a similar way as done in~\cite{Martin2020}.
The perturbations' power spectra will be obtained through an algorithm that computes the
valid interval for the adiabatic approximation and solves the dynamics of the
perturbations. With these results, we will compute the abundance of PBHs and compare our
results with other works in bouncing cosmologies.

This paper is divided as follows: Section~\ref{sec:bounce} is devoted to reviewing the
quantum bouncing model. In Sec.~\ref{linearpert} we define the quantized perturbations
around our background and discuss the Gauge problem. In Sec.~\ref{sec:formation} we
describe the general formation criteria for PBHs formed through critical collapse and
compute the critical threshold for the the bouncing scenario. In the same section, we
compute the PBH mass fraction and abundance in the bouncing model. We discuss and
conclude our results in Sec.~\ref{sec:discussion}.


\section{Flat-Dust Quantum Bouncing Background}
\label{sec:bounce}

In this section, we briefly discuss our adopted quantum bouncing background model, for
which we follow mostly Refs.~\cite{nelson_peter_bouncing_original, fluidgeral,
	nelson2021bouncing}. The primary motivation behind considering quantum bounces lies in
the necessity to evade the primordial singularity at the end of the contracting phase.
Classical General Relativity, as described by the Penrose-Hawking singularity theorems,
requires the violation of the null energy condition $\rho + p \geq 0$ to avoid the
singularity, a condition that often leads to instabilities in classical bouncing models.

Our quantum bouncing model is established through the canonical quantization of General
Relativity. The flat-dust bounce is thus a contracting universe model with a negative
effective energy density that dominates near the bounce, while classical behavior
prevails on larger scales. Nonetheless, the quantum contribution is sufficient to avoid
the singularity. This approach represents a conservative method in addressing quantum
gravity, as it applies the usual Dirac quantization techniques of constrained systems to
General Relativity \cite{nelsonhamiltonian}.

To apply canonical quantization to GR, we adopt the standard Hamiltonian formalism.  The
complete Wheeler-De Witt (WdW) equation that will arise from the Dirac quantization
poses several challenges that can be solved by assuming some additional hypothesis
~\cite{nelson2000bohm,halliwell1990introductory,nelson2021bouncing,nelson_bohm2023},
which we now discuss.  However, we should emphasize that the actual bouncing mechanism
is irrelevant to our results since the PBHs with relevant scales are formed way before
the quantum phase.

\subsection{Additional Hypothesis for Quantization}

The first problem arises because the WdW equation is formulated on superspace, which
represents the space of all possible metrics modulo diffeomorphisms and remains poorly
understood~\cite{halliwell1990introductory,dewitt1967}. Therefore, we shall only perform
quantization on a well-behaved sub-space that possesses the required symmetries, a
procedure that is known as a mini-superspace quantization. Thus, we quantize only the
sub-space of all possible flat FLRW geometries whose line element is given by the form
\begin{align}
	\label{physmetric}
	ds^2 = -N^{2}dt^2 + \bar{a}^2 \delta_{ij}dx^i dx^j
	,\end{align}
where $\bar{a}$ is the scale factor, $N$ is the lapse function and $\delta_{ij}$ is the
Kronecker delta. In this case, all information about the metric is stored in only one
degree of freedom, the scale factor $\bar{a}(t)$~\cite{nelson2021bouncing}.

Another problem is the lack of an explicit time evolution in the Hamiltonian of the
theory. This can be seen by attempting to write the conventional Schrödinger equation
for the total Hamiltonian and stating the lack of a clear time evolution parameter. This
fact is commonly referred to as the Problem of Time in Quantum Gravity ~\cite
{patrick_time_review,nelson_peter_bouncing_original,bianchi_time}. To define a
non-trivial propagator, we shall use an intrinsic variable of the system whose classical
evolution is monotonic. Subsequently, we require that the classical concept of time
emerges from this variable in the classical limit. Let us now apply such considerations
and discuss the solutions for this quantization procedure.

\subsection{Wheeler-DeWitt Equation Solutions}
\label{sec:wdw}

We start by applying the aforementioned considerations to a flat, homogeneous, and
isotropic universe containing a single perfect fluid characterized by its pressure
$\bar{p}$ and energy density $\bar{\rho}$, along with the equation of state $\bar{p} =
	w\bar{\rho}$, with a constant $w$. Subsequently, we specialize in the dust case where $w
	\approx 0$. In practice, we consider a barotropic fluid with a constant equation of
state, following the Schutz formalism as described in Ref.~\cite{fluidgeral}. In this
section we will briefly review the quantization of the flat-dust bouncing model,
focusing on the background dynamics.

To proceed with the quantization, we first derive the system's Hamiltonian from its
corresponding Lagrangian. The Lagrangian of the system is given by
\begin{align}\label{lagrangian}
	L & = \int \dd^3x \sqrt{-\bar{g}} \left[\frac{\bar{R}}{2\kappa}
		+ \bar{p}(\bar{v}, \bar{s})\right],
\end{align}
where, $\bar{g}$ denotes the determinant of the metric, $\bar{R}$ is the background
Ricci scalar, and $\kappa = \frac{8 \pi G}{c^4}$, where $c$ is the speed of light and
$G$ is the gravitational constant. We set $c = 1$ for simplicity. The term $\bar{p}$
represents the fluid pressure, which is expressed as a function of the specific enthalpy
$\bar{v}$ and specific entropy $\bar{s}$.

Using the Schutz formalism, the enthalpy can be decomposed into scalar potentials:
\begin{align}
	\bar{v}_\mu & = \bar\nabla_\mu \varphi_1
	+ \varphi_2 \bar\nabla_\mu \varphi_3
	+ \varphi_4 \bar\nabla_\mu \bar{s},
	\quad \bar{v} \equiv \sqrt{-\bar{v}_\mu \bar{v}^\mu},
\end{align}
where $\varphi_i$ are four independent scalar potentials. Varying the Lagrangian with
respect to these potentials and the specific entropy yields all thermodynamic relations,
including the perfect fluid energy-momentum tensor, entropy conservation, and particle
number conservation.

Since we assume a barotropic fluid with a constant equation of state $\bar{p} = w
	\bar{\rho}$, it follows that $\partial \bar{p} / \partial \bar{s} \vert_{\bar{\rho}}
	= 0$, where $\bar{\rho}$ is the energy density. This assumption simplifies the
pressure to the form $\bar{p} = f(\bar{s}) \bar{v}^{(1+w)/w}$.

After performing a Legendre transformation and solving the constraints, the Lagrangian
simplifies to:
\begin{align}
	L          & = \Pi_a \dot{\bar{a}} + \Pi_{\varphi_1} \dot{\varphi}_1 +
	N \left( \frac{\kappa \Pi_a^2}{12 V \bar{a}} - V \bar{a}^3 \bar{\rho} \right), \\
	\Pi_a      & = -\frac{6 V \bar{a} \dot{\bar{a}}}{N \kappa}, \qquad
	\Pi_{\varphi_1} = V \bar{a}^3 f \frac{1+w}{w}\bar{v}^{1/w},                    \\
	\bar{\rho} & = \frac{f}{w\bar{a}^{3(1+w)}} \left( \frac{w}{1+w}
	\frac{1}{Vf} \right)^{1+w}\Pi_{\varphi_1}^{1+w},
\end{align}
where $V$ is the conformal volume. Since the fluid term has the form
$\Pi_{\varphi_1}^{1+w}$ and the momentum term for the scale factor includes a factor
dependent on $\bar{a}$, we introduce a final canonical transformation to define new
momenta, $\Pi_q$ and $\Pi_T$
\begin{align}
	\Pi_q & = \Pi_{\bar{a}} \bar{a}^{(-1+3w)/2},                                                 \\
	\Pi_T & = \frac{V f}{w} \left( \frac{w}{1+w} \frac{1}{Vf}\right)^{1+w} \Pi_{\varphi_1}^{1+w}
\end{align}.

In the above framework, the total Hamiltonian takes the form
\begin{align}
	\label{total_hamiltonian_FLRW}
	{\cal H} & = \frac{N}{\bar{a}^{3w}}\dpar{\Pi_{T}- \frac{\kappa\Pi_q^{2} }{12}} ,
\end{align}
Canonical quantization is then performed by promoting classical variables to operators
satisfying the canonical commutation relations. This process yields the following
Wheeler-DeWitt equation for the wave-function of the universe  $\Psi(\bar{a}, T)$ (see
Ref.~\cite{nelson_peter_bouncing_original}):
\begin{equation}
	\label{wdweq}
	i\hbar\frac{\partial}{\partial T}\Psi(q,T) -
	\frac{\kappa\hbar^2}{12}\frac{\partial^{2}}{\partial q^{2}}\Psi(q,T)  = 0 ,
\end{equation}
such that a specific operator factor ordering was chosen to preserve the symmetries of
the classical system \cite{halliwell1990introductory}. Also, note that our last
canonical transformation leads to the following variable
\begin{equation}
	q = \frac{2\bar{a}^{\frac{3}{2}\dpar{1-w} } }{ 3\dpar{1-w } }.
\end{equation}
In the WdW equation above, we circumvent the problem of time by interpreting the
parameter $T$ as an intrinsic time variable that is monotonically related to the
classical cosmic time in the classical limit. This allows us to interpret the equation
as a Schrödinger-like equation.

Equation~\eqref{wdweq} resembles a time-reversed free particle Schrödinger equation and,
with appropriate boundary conditions, its solutions are wave functions of the scale
factor $\bar{a}$. We turn to the De Broglie-Bohm
interpretation~\cite{mukhanov2005physical, nelson2000bohm}, such that assuming a Gaussian
wave-function $\Psi(q, T)$, the Bohmian trajectory solution translates to the scale
factor reads
\begin{equation}
	\label{bohm_scale_factor}
	\bar{a}(T) = \bar{a}_{B}\dcol{1 + \dpar{ \frac{T}{T_{ B}} }^2 }^{{\frac{1}{3}\frac{1}{\dpar{1-w}} } }  ,
\end{equation}
where $\bar{a}_{B}$ is an integration constant that represents the minimum scale factor
value and $T_B$ is a small arbitrary constant related to the time scale of the bounce.
To simplify the analysis, we choose the lapse function $N = \bar{a}^{3w}$, but this
choice is only possible after solving the Bohmian trajectory. The reason is that,
initially, we cannot select $N$ as an operator function of the scale factor, However,
once we have the Bohmian trajectory $\bar{a}(T)$, we can make this choice, using it
purely for convenience in the analysis. In our model, $T$ extends from $T = -\infty$ in
the far past to $T = 0$ at the bounce, where the scale factor $\bar{a}$ reaches its
minimum. Remarkably, $\bar{a}(T) \neq 0$ for all $T$, which means that this model is
non-singular and represents an eternal universe~\cite{nelson_peter_bouncing_original}.
For a detailed derivation, see~\cite{nelson2021bouncing}.

It is important to clarify that in Ref.~\cite{nelson2021bouncing}, the author selects a
Gaussian wave function for the scale factor at the bounce. However, this does not imply
that this choice is the initial condition for our model. Generally, a Gaussian wave
packet of a one-dimensional free particle will develop a time-dependent phase
proportional to $q^2$. Thus, it is always possible to begin with a Gaussian wave
function at any time and define the bounce as the moment when this phase factor is zero.
For instance, in Ref.~\cite{Peter2016a}, the authors choose a Gaussian wave function at
any time and fix the phase using $H_0$.

Given that we focus solely on a contracting universe model filled with dust, we set $w
	\approx 0$, simplifying the time variable $T$ to the conventional cosmic time $t$. With
the scale factor obtained in Eq.~\eqref{bohm_scale_factor}, we derive its associated
Hubble function
\begin{equation}\label{bohm_hubble}
	\bar{H}(t) \equiv \frac{1}{\bar{a}}\frac{d\bar{a}}{dt} = \frac{2}{3}\frac{t}{ \dpar{ t^{2} + t^{2}_{b} } }\, ,
\end{equation}
and invert \eqref{bohm_scale_factor} to obtain
\begin{equation}\label{bohm_time}
	t(\bar{a}) = \pm t_{b}\sqrt{ \dpar{\frac{\bar{a}}{\bar{a}_{b}}}^{3} - 1 } .
\end{equation}

Eliminating the time $t$ using \eqref{bohm_time} in \eqref{bohm_hubble}, we find
\begin{equation}
	\label{friedman_scale}
	\bar{H}^{2} = \frac{ 4 }{ 9t^{2}_{b}}\dpar
	{ \frac{ \bar{a}^{3}_{b} }{ \bar{a}^{3} } - \frac{\bar{a}_{b}^{6}}{ \bar{a}^{6} } }\, ,
\end{equation}
which is equivalent to the Friedmann equations
\begin{equation}\label{effective_friedmann}
	\bar{H}^{2} = \frac{\kappa}{ 3 }\bar\rho - \bar{H}^{2}_{0}\Omega_{q0}\bar{a}^{-6},
\end{equation}
where $\bar\rho$ is the dust energy density. Also, from the derivative of
Eq.~\eqref{friedman_scale},
\begin{align}\label{effective_friedmann2}
	\dot{\bar{H}} = - \frac{\kappa}{2} \bar\rho + 3\bar{H}^{2}_{0}\Omega_{q0}\bar{a}^{-6}.
\end{align}

In the above equations, the overdot represents the time derivative and there is an
additional term $-\bar{H}^{2}_{0}\Omega_{q0}\bar{a}^{-6}$ when compared to the usual
Friedmann equation for a classical universe. In our model, the scale factor dynamics
resemble a typical Friedmann equation with a total energy density $\rho_{T} = \bar{\rho}
	+ \rho_{q},$ where $\bar{\rho}$ is the dust energy density, and $\rho_{q} = -\Omega
	\rho_{c} \bar{a}^{-6}$ denotes an effective negative energy density with an equation of
state $w_q = 1$ that accounts for quantum effects to avoid the primordial
singularity~\cite{vitenti2012large}. It is important to note that $\rho_{q}$ is not a
physical energy density but rather an effective one that emerges from the Bohmian
trajectory.

		 As a final remark, it is evident that, in the classical theory, one
		could interchangeably use either $\bar{H}^2$ or $\kappa \bar{\rho}/3$, as they are
		related by the classical Friedmann equation. However, in the quantum framework,
		$\bar{\rho}$ retains the same functional dependence on the scale factor, whereas
		$\bar{H}^2$ acquires an additional term due to quantum effects absent in the
		classical theory. Consequently, when deriving the Lagrangian for perturbations, it
		is crucial to avoid relying on the classical Friedmann equations. This approach was
		explicitly followed in Ref.~\cite{fluidgeral}, where the authors computed
		perturbations up to the second order without invoking any classical background
		equations.

		To compute perturbations in a Bohmian trajectory, one employs
		Eqs.~\eqref{effective_friedmann} and \eqref{effective_friedmann2} to determine the
		Hubble function and its time derivative, while the energy density $\bar{\rho}
			\propto \bar{a}^{-3}$ (with $\bar{a}$ given by Eq.~\eqref{bohm_scale_factor}) is
		used for calculating energy density perturbations. This ensures, for instance, that
		terms such as $1/(\bar{\rho} + \bar{p})$ remain well-defined even when $\bar{H}^2$
		or $\dot{\bar{H}}$ pass through zero.

With this, we conclude the analysis of our background quantum bouncing model and move to
its associated perturbations. In the next sections, we will consider the above quantum
bouncing model with $w\approx 10^{-10}$ and $\bar{a}(t_0)/a_B = 10^{35}$ where $t_0$ is
the time during the contracting phase where $H(t_0) = -H_0$ for $H_0 = 70 \,
	\text{km/s/Mpc}$. Moreover, when solving the equations numerically, we use the exact
expressions in terms of $w$ without approximation.


\section{Linear Scalar Perturbations}
\label{linearpert}

While the universe appears homogeneous and isotropic on large scales, the FLRW metric
falls short of providing a precise description of our universe, which presents
inhomogeneities that are associated with structure formation, e.g. galaxy clusters,
black holes, stars, and others. To characterize the physical universe, we study linear
scalar perturbations around our flat contracting background metric described in the
previous section. We will then investigate how said perturbations may seed the formation
of primordial black holes in the subsequent section. We will from now on characterize
background quantities with an overbar and perturbed (physical) quantities without it.

\subsection{Gauge Invariant Perturbations}

We follow mainly the perturbation theory developed in Refs.~\cite{covariant_bardeen,
	Vitenti2012}. Assuming only scalar perturbations, the total metric of the physical
space-time is given by
\begin{align}
	\label{physmetric-pert}
	\dd s^2 = -(1- 2 \phi)\dd t^2 + \bar{a}\bar{D}_i \mathcal{B} \dd t \dd x^i +
	\bar{a}^2(1-2 \psi)\delta_{ij}\dd x^i \dd x^j - \bar{D}_i\bar{D}_j \mathcal{E} \dd x^i \dd x^j
	.\end{align}
Here, $\phi, \psi, \mathcal{B}$ and $\mathcal{E}$ denote the scalar metric perturbations
that we will assume to be much smaller than one \cite{vitenti2012large}. Also,
$\bar{D}_i$ is the spatial covariant derivative in the $i$-th direction. Note that our
barotropic fluid has no anisotropic pressure.

In the perturbative treatment, we continue to use the Schutz formalism and construct the
Lagrangian as in Eq.~\eqref{lagrangian}, but now incorporating the perturbed metric. The
perturbed Lagrangian and associated variables are defined as in Ref.~\cite{Vitenti2012}.
Note that in the perturbed Lagrangian, the perturbations are always linked to the
physical quantities and their corresponding background values. The effective energy
density and pressure do not fluctuate; instead, the quantum effects from the background
are incorporated through the use of the background Bohmian trajectories.

The metric perturbations are gauge-dependent variables. In cosmology, a gauge can be
seen as the freedom in how we connect, or map, the background and physical manifold, and
how we choose our coordinate system~\cite{vitenti2012large}. Since GR is a covariant
theory, this freedom may then be interpreted as a gauge, which may lead to an ambiguous
description of perturbations. Depending on the foliation that characterizes the manifold
hyper-surfaces and how we define the perturbations around our background, the physical
quantities may have different values \cite{covariant_bardeen}. Thus, it is recommended
that we work with gauge independent variables to carry out our computations and go back
to the physical variables at the end when necessary \cite{mukhanov2005physical}.

We define gauge invariant quantities by analyzing their transformations under gauge
transformations, such that we can combine different gauge-dependent quantities to form
new invariant ones~\cite{covariant_bardeen}. However, this leads to a freedom on the
variable definitions since many combinations of variables may lead to gauge invariant
quantities. With this in mind, we define the Bardeen gauge invariant
variables~\cite{Bardeen1980}
\begin{align}
	\label{bardeen1}
	\Phi & \equiv \phi + \dot{\delta \sigma}\, , \\
	\label{bardeen2}
	\Psi & \equiv \psi - \Bar{H}\delta \sigma
	,\end{align}
with
\begin{align}
	\delta\sigma = -(\dot{\mathcal{E}} - \mathcal{B}) + {2 \Bar{H}\mathcal{E}}
	.\end{align}
It is important to notice that the new variables in
Eqs.~\eqref{bardeen1}-\eqref{bardeen2}, and other gauge invariant variables do not have
a physical meaning unless a gauge is chosen. For instance, in the case of the Newtonian
gauge, $\delta\sigma = 0$ and $\Phi$ represents the Newtonian potential. Hence we must
define our gauge invariant quantities such that they represent our desired physical
variables when we assume a particular gauge choice.

We are mostly interested in the energy density perturbations that collapse to form PBHs.
These perturbations can be examined using the density contrast, defined as
\begin{align}
	\label{densitycon}
	\delta & \equiv \frac{\delta \rho}{\bar{\rho} + \bar{p}},
\end{align}
where $\delta\rho$ is the perturbation to the background energy density $\bar\rho$.
	 Note that the density contrast defined above appears in this form
		when the second-order perturbations Lagrangian is derived. This means
		that as long as $\bar{\rho} + \bar{p} > 0$, the density contrast is well-defined.
		As discussed in Sec.~\ref{sec:wdw}, when computing the perturbations,
		we use the Bohmian trajectory to determine the background quantities,
		which ensures that $\bar{\rho} + \bar{p} > 0$ for all times. 

The density contrast provides a normalized measurement of the energy density
perturbation around the background matter density field. However, once again we are
interested in its gauge-invariant form
\begin{align}
	\label{deltarhoinvariant}
	\tilde{\delta\rho} & \equiv \delta\rho - \mathcal{V} \dot{\Bar{\rho}}
	,\end{align}
where $\mathcal{V}$ is the velocity perturbation of the fluid, which has its gauge
invariant form
\begin{align}
	\tilde{\mathcal{V}} & \equiv \mathcal{V} + \delta\sigma
	.\end{align}

Using the background Einstein equations, we can relate the density contrast to the
Bardeen variables and the gauge invariant curvature perturbation $\zeta$ through the
following relations~\cite{Mukhanov1992, Vitenti2013}
\begin{align}
	\Psi                                                    & =\Phi,
	\\
	\label{deltarho}
	-\frac{2\bar{D}^2 \Phi}{3  \kappa (\bar{\rho}+\bar{p})} & =   \frac{{\tilde{\delta\rho}}} {3(\bar{\rho}+\bar{p})}
	,
	\\
	\label{vrelation}
	\zeta                                                   & \equiv \Psi  + \bar{H} \tilde{\mathcal{V}}
	\\
	\label{zeta2}
	\zeta                                                   & = \frac{3  \bar{a}^3}{N^2z^2 c_s^2 } \left[\frac{\partial}{\partial t}\left(\frac{\Phi }{3\bar{H}}\right) + \frac{\Phi}{3}\right]
	,                                                                                                                                                                                           \\
	\label{phiz}
	\bar{D}^{2} \Phi                                        & =  - \frac{z^2\bar{H}}{\bar{a}^3} \dot{\zeta} = -\frac{\bar{H}}{2\bar{a}^3} \Pi_\zeta
	.\end{align}
In the above, $c^{2}_s=\dpar{ \frac{\partial\bar{\rho}}{\partial\bar{p}} }_{S}= w$ is
the speed of sound, $\Pi_\zeta$ is the conjugated momenta related to $\zeta$ and
\begin{align}
	\label{zdef}
	z^2=\frac{\kappa \bar{a}^3 (\bar{\rho} + \bar{p})}{2\bar{H}^2 c_{s}^2}.
\end{align}
For completeness, we also write the relation between the curvature perturbation and the
usual Mukhanov-Sasaki variable $v$, that is,
\begin{align}
	\label{msv}
	v & \equiv - \zeta z \sqrt{\frac{2}{\kappa}}
	.\end{align}
In the next section, we will establish a connection between PBH formation and density
contrast. For now, it suffices to understand that the excess energy density associated
with the perturbations leads to black hole formation, and we can measure such excess
through the density contrast. Thus our definition in Eq.~\eqref{deltarhoinvariant} is
extremely important and we will now analyze it.

Note from Eq.~\eqref{deltarhoinvariant} that if we choose a gauge where $\mathcal{V}
	=0$, $\Tilde{\delta\rho}$ becomes the physical density perturbation, which leads to an
easier interpretation of this quantity. Also, the choice of $\mathcal{V} =0$ will be
well suited to connect our perturbation theory with the Lemaitre-Toman-Bondi metric
discussed in App.~\ref{appgauge}.  Other works have defined the gauge invariant density
perturbation as
\begin{align}
	\label{newtoniandeltarho}
	\Tilde{\delta\rho}^{N} & \equiv \delta\rho - \delta\sigma \dot{\Bar{\rho}}
	,\end{align}
so that this quantity becomes the physical variable in the Newtonian Gauge, where
$\delta\sigma = 0$ and $\Phi = \phi$. However, in Ref.~\cite{vitenti2012large}, it was
shown that for bounce models with long contracting phases, regardless of the bounce
type, the Bardeen perturbation $\Phi$ grows larger than one and invalidates the
perturbative series, as $\phi$ also grows larger than one in the Newtonian Gauge.
Additionally, the definition in Eq.~\eqref{newtoniandeltarho} leads to
\begin{align}
	\zeta & = \Phi + \frac{2\bar{D}^2 \Phi}{3  \kappa (\bar{\rho}+\bar{p})} +
	\frac{{\tilde{\delta\rho}^N}} {3(\bar{\rho}+\bar{p})},
\end{align}
which implies that $\tilde{\delta\rho}^N$ grows with $\Phi$ and thus has larger spectra
as well as in this gauge. Hence, the Newtonian gauge is not a valid choice for bounce
models as it would lead to miss-calculation of the physical quantities whose values
would be inflated. We shall avoid this choice in this work and stick with the definition
in Eq.~\eqref{deltarhoinvariant}. We now compute the density contrast modes, which
require the perturbative equations of motion of the model.

\subsection{Classical Equations of Motion}

The perturbations are described by the Einstein-Hilbert action expanded up to second
order, resulting in the Mukhanov-Sasaki Lagrangian
\begin{align}
	\label{lagms}
	L_{MS} & = \int \dd^3\textbf{x} \frac{1}{ \kappa} \left({\dot{\zeta}}^2 z^2 + c_s^2 z^2 \zeta\Delta \zeta\right)
	,\end{align}
where $\Delta = \bar{D}^2$ is the spatial Laplacian operator and $z$ is given by
Eq.~\eqref{zdef}. To compute the density contrast in Eq.~\eqref{deltarho}, we will need
the curvature modes related to the Mukhanov-Sasaki Lagrangian. The extremization of
Eq.~\eqref{lagms} in Fourier space yields the equation of motion
\begin{align}
	\label{four}
	\Ddot{\zeta}_{k}+ 2\frac{ \dot{z} }{ z } \dot{\zeta}_k + \frac{k^2}{\bar{a}^2} \zeta_k & =0
	,\end{align}
where $k$ is the comoving wave number of the modes. Since obtaining analytical solutions
to \eqref{four} is non-trivial, we will employ a numerical code to compute the modes,
which requires the use of dimensionless variables. In dimensionless units, denoted with
a subscript $A$, we redefine our variables as\footnote{A derivative of a dimensionless
	variable will also be dimensionless. However, we use the same dot notation.}
\begin{align}
	\zeta_{k_A, A}   & \equiv \frac{\zeta_{k}}{\sqrt{ \kappa \hbar R_H}},           \\
	\Pi_{\zeta_k, A} & \equiv \frac{\Pi_{\zeta_k} \sqrt{R_H}}{\sqrt{\kappa \hbar}}, \\
	\label{kadm}
	k_A              & \equiv k R_H,                                                \\
	t_A              & \equiv \frac{t}{R_H}
	,\end{align}
where $R_H = \frac{1}{\bar{H}_0}$. The equation of motion becomes
\begin{align}
	\label{eomosc}
	\dot{\Pi}_{k_A, A}+  \frac{2 c^2_s z^2 k_A^2}{\bar{a}^2  } \zeta_{k_A, A} & =0 \nonumber~~ or \\
	\dot{\Pi}_{k_A, A}+  2 \omega_{k_a}^2 z^2  \zeta_{k_A, A}                 & =0
	.\end{align}
Here, $\omega_A^2 \equiv \frac{c_s^2 k_a^2}{\bar{a}^2}$ is the dimensionless frequency
and the dimensionless conjugated momenta are given by
\begin{align}
	\Pi_{k_A, A} & = 2 z^2 \dot{\zeta}_{k_A, A}
	.\end{align}
Eq.~\eqref{eomosc} can be interpreted as a harmonic oscillator with a time-dependent
frequency $\omega_{k_a}(t_a)$ and mass $m(t_a) = 2z^2$. From here on, we are going to
use the dimensionless variables with the same notation as described above.

Before solving this system numerically, we need to discuss the quantized version of our
modes. This is because we need a prescription to set initial conditions of the
perturbative variable $\zeta$. We do that by imposing initial vacuum conditions for the
quantum fields, and thus, the problem turns into the problem of defining an appropriate
vacuum state.

\subsection{Quantization}

In this model, the background is characterized by a quantum contracting phase where an
effective quantum fluid dominates near the bounce. Consequently, we seek quantized
perturbations for consistency, although both theories can be viewed independently. In
particular, the use of quantized perturbations evolving on classical backgrounds has
been widely used since the results of Mukhanov and Chibisov \cite{mukhanov1981quantum}
and Hawking \cite{hawking1982quantum_fluctuations} to derive the power spectrum of
primordial perturbations, which are in turn used to describe the formation of structure
in our universe. In our model, we quantize both the background and the
perturbations. Since we adopt the De Broglie-Bohm interpretation for the background,
quantum effects are described in terms of the Bohmian scale factor, as shown in
Eq.~\eqref{bohm_scale_factor}. This allows us to treat quantum perturbations evolving on
this scale factor. To demonstrate the consistency of this approach,
Ref.~\cite{Peter2005} derived the second-order Lagrangian for tensor perturbations
without assuming classical equations of motion for the background. Furthermore, by using
a factorized wave function for the perturbations and the background, the authors showed
that these perturbations evolve conditioned on the Bohmian scale factor. In
\cite{Vitenti2013}, the equivalent Lagrangian for scalar fluid perturbations was
derived and is the one we use in this work.

The use of quantum fields to describe primordial perturbations means that said fields
will have statistical properties. We may then partition the universe into local spatial
regions and consider each one as a realization of a random process and compare its
statistical properties with our theoretical predictions.\footnote{Here, an important
	remark must be made: in inflationary models, one usually postulates this quantum to
	classical statistical connection, but the specific mechanism that converts a quantum
	universe to a classical one is still an open problem
	\cite{nelson2021bouncing,mukhanov2005physical}. In our model, even if the perturbative
	level predictions do not depend strongly on the bouncing mechanism if one assumes the De
	Broglie-Bohm interpretation applied to Canonical Quantum Gravity, this problem is
	automatically solved \cite{nelson_bohm2023}. }

We proceed to quantize our fields by promoting them to Hermitian operators that act on a
Fock space. In terms of the usual Fourier mode expansion, our quantum operators, denoted
with a hat superscript~$\hat{ }$~from now on, are given by
\begin{align}
	\label{vexp}
	\hat{\zeta}(\textbf{x}, t) & = \frac{1}{(2\pi)^{\frac{3}{2}}}\int \mathrm{~d}^3{\textbf{k}} \left(e^{i{\textbf{k}}.\textbf{x}}\zeta_{k}^{*}(t) a_k + e^{-i{\textbf{k}}.\textbf{x}}\zeta_{k}(t) a^{\dagger}_k\right)
\end{align}
and
\begin{align}
	\label{pvexp}
	\hat{\Pi}_\zeta(\textbf{x}, t) & = \frac{1}{(2\pi)^{\frac{3}{2}}}\int \mathrm{~d}^3{\textbf{k}} \left(e^{i{\textbf{k}} .\textbf{x}}\Pi_{\zeta_k}^{*}(t) a_k + e^{-i{\textbf{k}} .\textbf{x}}\Pi_{\zeta_k}(t) a^{\dagger}_k\right)
	,\end{align}
where $\textbf{k}$ is the momentum vector with modulus $k$ and $a_k$ and $a_k^{\dagger}$
are the annihilation and creation operators respectively. Demanding that the quantum
fields satisfy the canonical commutation relations
\begin{align}\label{commutation_relations}
	\left[\hat{\zeta}(\textbf{x}, t) , \hat{\Pi}_\zeta(\textbf{y}, t) \right] = i\hbar\delta(\textbf{x} - \textbf{y})\, ,
\end{align}
and that the complex modes satisfy
\begin{align}
	\label{basec}
	\dot{\zeta}_{k} \zeta_{k}^{*} - \zeta_{k}\dot{\zeta}_{k}^{*} & =i\hbar\, , \\
	\dot{\Pi}_{k} \Pi_{k}^{*} - \Pi_{k}\dot{\Pi}_{k}^{*}         & =i\hbar\, ,
	.\end{align}
imply that the creation and annihilation operators $ a_k,a_k^{\dagger} $ satisfy the
usual creation and annihilation algebra
\begin{align}
	\left[{a}_{k_1}, {a}_{k_2}\right] & = \left[{a}^\dagger_{k_1}, {a}^\dagger_{k_2}\right] = 0
\end{align}
and
\begin{align}
	\label{acom}
	\left[{a}_{k_1}, {a}^\dagger_{k_2}\right] = \delta(k_1 - k_2)
	,\end{align}
where $\delta(k_1 - k_2)$ is the dirac delta between the momenta.

Equation~\eqref{basec} represents the only constraint for choosing the basis of our
problem and it is known as the vacuum normalization. To fully specify our operators and
solve Eq.~\eqref{eomosc}, we need to set initial conditions that will physically
determine the annihilation operator and, consequently, the vacuum state of the theory.

\subsection{Adiabatic Vacuum}

In quantum field theory, due to the non-applicability of the Stone-Von Neumann Theorem,
different choices of Hilbert Space that are consistent with the commutation relations in
Eq.~\eqref{commutation_relations} are not unitary equivalent, which means that they lead
to different physical predictions \cite{wald1994quantum}. Therefore, one also needs a
prescription to construct the associated Hilbert Space of a quantum field theory.

Although in usual Minkowski space-time one may use its symmetries to define the Hilbert
space, in curved spacetimes one needs other techniques to construct said
space~\cite{birrell1984quantum}. This problem can be mapped to a choice of operators
$\hat{a}_{k}$ that annihilate the vacuum state $\ket{0}$, which in turn depends on the
choice of mode functions $\zeta_{k}(t)$ that satisfy the normalization condition in
Eq.~\eqref{basec}~\cite{mukhanov2007introduction}. Since such a condition is preserved
by the time evolution, it suffices to choose a set of initial conditions
$\dcha{\zeta_{k}(t_{0}), \Pi_{k}(t_{0})}$ at an initial time $t_{0}$ \cite{vacuum2022}.

A widely used prescription to define such a vacuum state is known as the adiabatic
vacuum prescription~\cite{birrell1984quantum, mukhanov2007introduction}. Its main idea is
to set initial conditions by demanding that the mode functions $\zeta_{k}(t)$ coincide
with their ${\cal N}$ order adiabatic approximation, $^{({\cal N})}\zeta_{k}(t)$. This
is implemented if one chooses the initial conditions~\cite{birrell1984quantum}
\begin{align}
	\label{v_init}
	{ }^{(\mathcal{N})}	\zeta_k(t_0) & = \frac{1}{\sqrt{m(t_0)\omega_k}(t_0)}e^{i\alpha_k(t_0)}~\text{and}~ { }^{(\mathcal{N})}	\dot{\zeta}_k(t_0) = im(t_0)\omega_k(t_0)\zeta_k(t_0),
\end{align}
where $\omega_k^2(t) = \frac{c_s^2 k^2}{\bar{a}^2}$, $m(t) = 2z^2$ and $\alpha_k$ is
defined by the relation
\begin{align}
	{ }^{(\mathcal{N})}	\zeta_k(t_0) \dot{\alpha}_k(t_0) & = 1.
\end{align}

Using these initial conditions for the modes, a well-defined vacuum state is obtained,
which is in turn used to construct the Fock Space by successive applications of the
creation operator $\hat{a}^{\dagger}_{k}$ and their linear combinations
\cite{mukhanov2007introduction}. Now that we have defined the appropriate initial
conditions for quantization, we may use them to solve the equations of motion in
Eq.~\eqref{eomosc} and obtain the mode functions, which we shall do numerically.
Finally, it is important to note that the adiabatic vacuum prescription is consistent
with the perturbative analysis of the model, as it leads to small perturbations as
discussed in Ref.~\cite{vitenti2012large}.

\subsection{Numerical Solution}

The equations of motion for the curvature perturbation modes are solved numerically
using the Numerical Cosmology library~(NumCosmo)~\cite{Vitenti2014}. Specifically, the
NcmCSQ1D and the NcHIPertAdiab algorithms are employed for this purpose. The equations
are split into two parts: one representing a harmonic oscillator with a mass $m_A$ and
the other representing the time evolution of the conjugate momenta $\Pi_{k_A, A}$ of the
modes. Explicitly,
\begin{align}
	\Pi_{k_A, A} & = m_A \dot{\zeta}_{k_A, A}
\end{align}
and
\begin{align}
	\dot{\Pi}_{k_A, A} = -m_A\omega^2_A \zeta_{k_A, A}
	,\end{align}
The adiabatic vacuum prescription is considered, and the initial conditions for the
modes are set according to Eqs.~\eqref{v_init} up to the fourth order\footnote{This is
	because the adiabatic approximation leads to an asymptotic series, whose precision drops
	as one goes up to a certain definite order~\cite{mukhanov2007introduction}. In
	particular, the code determines the order of optimal precision.}. These initial
conditions are used as inputs for the numerical algorithm to compute the Fourier modes
of the curvature perturbation modes ${\zeta}_{k_A, A}$ and their conjugate momenta
$\Pi_{k_A, A}$. Also, the numerical code computes the validity of the adiabatic
approximation for different intervals of time. To assure the veracity of the code, the
algorithms have been validated with unit testing and we compared them with analytical
solutions (see
\href{https://github.com/NumCosmo/NumCosmo/blob/master/tests/test_py_hipert_adiab.py}{GitHub})\footnote{Also,
	see the
	\href{https://github.com/NumCosmo/NumCosmo/blob/master/notebooks/primordial_perturbations/single_fluid_qb.ipynb}{Jupyter
		Notebook} for an example on the usage of the code.}


\subsection{Spectra}
\label{sec:spectra}
\begin{figure}[tbp]
	\centering
	\includegraphics[width=.6\textwidth]{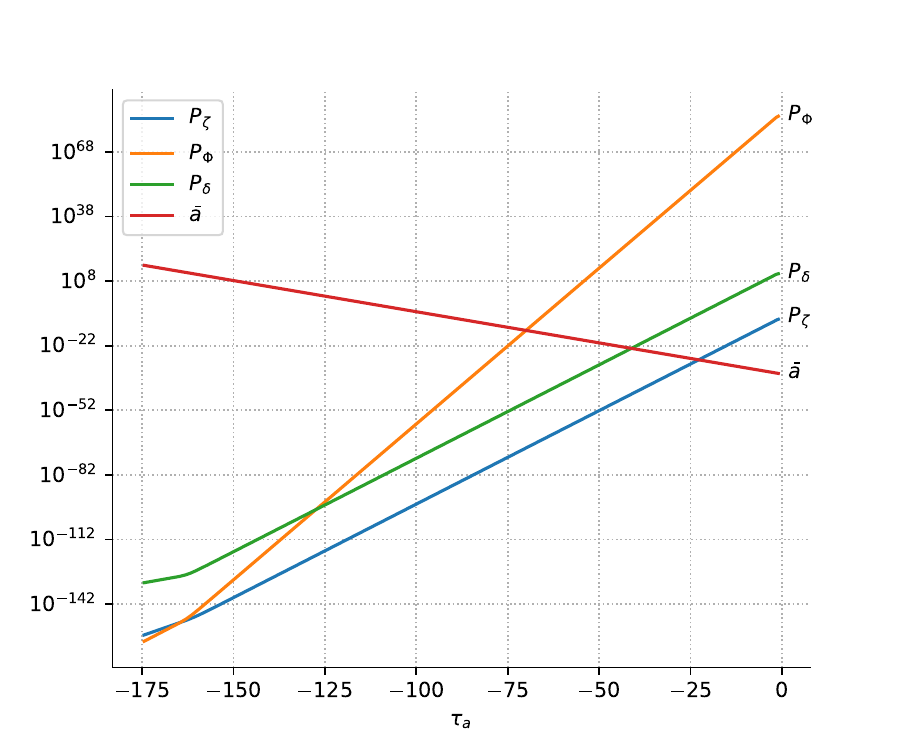}
	\caption{Plot of the power spectrum of the curvature perturbation modes $\zeta_k$,
		the Bardeen field variable $\Phi_k$ and the density contrast $\delta$ computed with
		the NumCosmo library for $w = 10^{-10}$ and $x_b = 10^{35}$. The red plot represents
		the scale factor evolution in time. The time parameter in the x-axis is given by
		Eq.~\eqref{tauparameter}.}
	\label{fig:figspec}
\end{figure}

The power spectrum of a theory plays a crucial role in understanding the formation of
large-scale structures. The power spectrum also allows one to completely describe the
statistics of the problem if the density field is described by Gaussian
fluctuations~\cite{Baugh}. To compute the power spectrum in our cosmological model, we
first calculate the two-point correlation function of the field variable
$\zeta(\textbf{x},t_i)$ at an initial time $t_i$. This function measures the spatial
correlation between fluctuations at different points in space. The correlation function
is expressed as an integral over Fourier modes, yielding the desired spatial
correlation,  and takes the form
\begin{align}
	\left<\hat{\zeta}(\textbf{x},t_i)\hat{\zeta}(\textbf{y},t_i)\right>
	 & =\frac{1}{(2\pi)^\frac{2}{3}}\int \mathrm{~d}^3{\textbf{k}}\left[ |\zeta_{k}(t_i)|^2 e^{-i\textbf{k}(\textbf{x} - \textbf{y})}\right] \nonumber \\
	 & =\int\frac{ \mathrm{~d}{k}}{k}\left[ P_{\zeta}(k)  \frac{\sin{kR}}{kR} \right]
	,\end{align}
where $R = \abs{\textbf{x} - \textbf{y}}$ and we performed an integral over solid angles
in the last equality. In the above, the power spectrum $P_{{\zeta}}$ is defined
as\footnote{Note that the power spectrum is already dimensionless, and thus we can use
	the same definition with our dimensionless variables as well.}
\begin{align}
	\label{eqspec}
	P_{\zeta}(k) & \equiv \frac{k^3|\zeta_k(t_i)|^2}{2\pi^2}
	.\end{align}
Using Eq.~\eqref{eqspec} and the numerical code described in the last section, we can
plot the spectra of the theory in Fig.~\ref{fig:figspec}. Once again we have different
initial times for the adiabatic limit for distinct modes. We also notice an increasing
spectra for all the modes, such that they peak at the bounce time.

As known from the literature, the dust bouncing model has an approximate scale-invariant
spectrum, that is
\begin{align}
	\label{powerlaw}
	P_{{\zeta}}(k) & \approx A k^{n_s-1}
	,\end{align}
where $A$ is a constant and $n_s$ is the spectral index. This is a general feature of
single barotropic fluid quantum bouncing models, and the spectral index is related to
the equation of state parameter $w$ by\footnote{For an explicit semi-analytic
	derivation, see \cite{nelson_peter_bouncing_original}.}
\begin{equation}
	n_{s}(w) = 1 + \frac{ 12w }{ 1 + 3w }\, ,
\end{equation}
which is nearly scale invariant for $|w| \ll 1 $. However, the usual positive values of
$w$ lead to a blue tilted spectrum $n_{s} > 1$, which differ from CMB observations
\cite{planck_inflation_constraints}, consistent with the inflationary scenario
prediction of a red tilted spectrum with $n_{s} \approx 0.96$. Although the initial
power spectrum is not consistent with observations, one must recall that our model is
considering a pure dark matter-dominated universe, neglecting the effects of radiation.
In particular, it has been suggested \cite{nelson2021bouncing} that the inclusion of
radiation may lead to a red almost scale-invariant spectrum. Therefore, this model must
be understood as a first approximation to a more complete model, which is still under
development. Furthermore, since most of the modes that influence our universe have
crossed the Hubble horizon during dust domination, this means that this model, even with
its simplicity, may still cover relevant information for the future complete model.

This result should not advocate for the prediction failure of the contracting scenario
since some works also suggest a blue-tilted spectrum for inflationary
models~\cite{Wang2014, Cai2015}. For instance, in Ref.~\cite{Kuroyanagi2021} they
discuss the possibility of explaining the recent NANOGrav results with a blue-tilted
spectra~\cite{Wu2023}. Also, the spectra are only dictated by a power law for narrow
intervals of momenta, which implies that corrections to Eq.~\eqref{powerlaw} must be
applied. Nonetheless, we see that the contracting phase produces a nearly
scale-invariant spectrum. In the next section, we begin to analyze the primordial black
hole formation seeded by the scalar perturbations in our cosmological background.

The resulting plots in Fig.~\ref{fig:figspec} depict the power spectra for a single
momentum mode ($k_a = 0.1$) of the curvature perturbation, the Bardeen field variable,
the density contrast, and the evolution of the scale factor over time. For visualization
purposes, we use the same time parameter as the one in the numerical code, namely,
\begin{align}
	\label{tauparameter}
	\cosh(\tau_a)^{2} & \equiv \left[1 + \left(\frac{t}{t_b}\right)^2\right]
	,\end{align}
such that $\tau_a$ is dimensionless by definition. The plots show an increasing power
spectrum for all fields, peaking at the bounce. Note that $\Phi$ has highly divergent
spectra, as discussed previously,  which shows the inapplicability of the Newtonian
Gauge in this model. Additionally, the scale factor decreases with time until reaching
its minimum value at the bounce.

\newpage

\section{PBH formation during the contracting phase}
\label{sec:formation}

Various mechanisms can result in Primordial black hole formation (see
Ref.~\cite{Escriva2023} for a review). In this work, we focus on investigating the
formation of primordial black holes through the critical collapse of matter
perturbations. This approach has been extensively studied since its popularization by
Carr and Hawking~\cite{Hawking1971, Carr1974}, and it has found broad applications in
the context of inflationary models. We want to extend this approach to the context of a
quantum dust-bouncing model and compute the PBH abundance in this scenario. It is
important to note that although we perform all the calculations with the gauge invariant
variables, we will analyze our results for the physical density perturbation in the
fluid's gauge, as defined in App.~\ref{appgauge}.

Before we proceed, we must emphasize that we are considering the formation
of PBHs in a classical context. We are using the quantum fields to describe the initial
conditions of the perturbations, but the contracting phase is amplifying their
amplitudes in a similar fashion to an inflationary phase. Such assumptions, particularly
regarding the quantum-to-classical transition, are commonplace in the literature,
especially in inflationary models. For example, in Ref.~\cite{Q2C2}, the authors discuss
the quantum-to-classical transition in the context of the Bohmian-De Broglie
interpretation, showing that the quantum potential is negligible when the perturbations
are amplified by the inflationary phase.

In this work, we are not discussing the quantum-to-classical transition but are assuming
that the perturbations are classical when amplified by the contracting phase. Moreover,
when approaching the bounce time, the background itself is not a classical solution but
a quantum one. Thus, in order to study the collapse of the perturbations at this phase,
one must consider the quantum effects of the background. However, it is important to
note that PBHs with masses within the observable ranges are formed during the
contracting phase, where it is safe to assume that the initial quantum fields have
already transitioned to classical behavior and are far removed from the bounce phase.

\subsection{Formation Criteria}

In our context, critical collapse is the collapse of matter perturbations when they
achieve a given threshold. The energy-density perturbation is characterized by the
density contrast $\delta$ defined in Eq.~\eqref{densitycon}. We assume that if $\delta >
	\delta_c$, where $\delta_c$ is the critical threshold, the perturbations will collapse
into a black hole. This critical threshold relies both on the cosmological model and the
shape of perturbations (see Ref.~\cite{Niemeyer1998, Musco2019} for a comprehensive
analysis of this effect) and therefore must be carefully studied as it will heavily
impact the formation of black holes.

The critical collapse model assumes the existence of a region of radius $r$ with an
overdensity $\delta$. To compute the probability of the existence of this overdense
region, we will need to compute the variance of our random variable $\delta$ and instead
of working with the original density contrast, we work with its filtered version
$\delta_r$, which is defined by
\begin{align}
	\delta_r(t) & =  \int d^{3}\textbf{x}^\prime \delta(t, \vec{x}^\prime) W_r\left(\vec{x} - \vec{x^\prime} \right)
	,\end{align}
where $r$ is a smoothing scale in comoving units related to the mean density by $M =
	\frac{4\pi r^3 \bar{\rho}}{3}$ and $W$ is the top-hat filter
\begin{align}
	\label{windowf}
	W_r(\vec{x}) & =W_r\left(|\vec{x}| = x\right) =
	\begin{cases}
		\frac{1}{4\pi r^3} & \text{if $x \leq r$} \\
		0                  & \text{otherwise}
	\end{cases}
	.\end{align}
The filter function is used to select a desired scale such that we would only analyze
collapsed objects that lie in this interval\footnote{Frequently we will usually refer to
	such scale in terms of the mass $M$ instead of the comoving radius $r$ that encloses
	this mass.}. In this work, we are using a simple top-hat windowed function for the
filter. We usually work on the Fourier space with comoving wave-number $k$, such that
the filter is given by
\begin{align}
	W_r(k) & = \frac{3}{(kr)^2}\left(\frac{\sin(kr)}{kr} - \cos(kr)\right) = \frac{3}{kr}j_1(kr),
\end{align}
where $j_1$ is the spherical Bessel function of the first kind. We will see that the
filter is necessary to analyze scales where the PBH formation is most likely to happen.

Coming back to $\delta_r$, we know from our quantum fields that the density contrast is
a random field following a Gaussian distribution with a zero mean. This is also
supported by significant research on the statistics of peaks in random Gaussian fields,
which can give rise to collapsed objects, as explored in
Ref.~\cite{Bardeen1986statistics}. Mathematically, the probability of a region with
radius $r$ having an overdensity $\delta_i$ is given by
\begin{align}
	\label{gaussdelta}
	P(\delta_i) & = \frac{1}{\sqrt{2\pi} \sigma_r} \exp(-\frac{\delta^2_i}{2 \sigma_r^2}).
\end{align}
This Gaussian is completely defined by its variance
\begin{align}
	\label{eqsigmar}
	\sigma_r^2 & =\left< \delta_r^2(\vec{x}) \right> \equiv \sigma_r^2 = \frac{1}{2\pi^2} \int_0^\infty \frac{ \mathrm{~d}{k}}{k}\left[ P_{\tilde{\delta}}(k) \abs{W_r(k)}^2\right],
\end{align}
where $P_{\tilde{\delta}}(k)$ is given by Eq.~\eqref{eqspec} with the density contrast
modes. Unfortunately, for our bounce model, this integral does not converge. Before we
perform this calculation, we will analyze the critical threshold for PBH formation,
which will lead to scale constraints that will help us solve this problem.

\subsection{Critical Threshold}

\label{critical_delta}
In inflationary models, the frozen super-Hubble density perturbations re-enter the
Hubble horizon at the end of the potential decay and collapse into a black hole if they
have values above the critical threshold~\cite{Martin2014}. Hence the Hubble horizon is
viewed as a characteristic scale for their formation and the threshold ($\delta_c$) is
obtained via the Misner-Sharp equations (see Ref.~\cite{Musco2019}). However, in the
bounce scenario, the perturbations constantly evolve in time and there is not only one
characteristic scale for the formation, since any perturbation above a critical
threshold may collapse as they are not frozen. Thus, we must carefully analyze how to
obtain this critical value starting with a local metric for the collapse.

	It is important to note that we are assuming the perturbations are
		classical at this stage. Otherwise, it would be necessary to consider a quantum
		model for the collapse, which would require developing a quantum Misner-Sharp
		framework with initial conditions derived from the quantum fields. This is a highly
		complex problem that remains largely unexplored in the literature.

\begin{figure}[tbp]
	\centering
	\includegraphics[width=.8\textwidth]{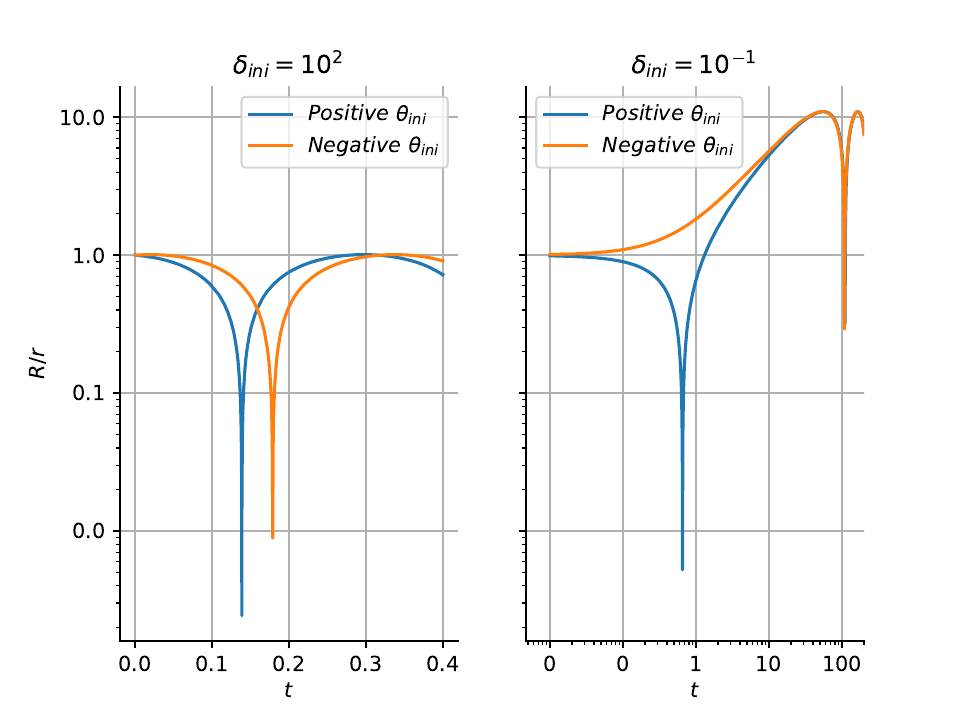}
	\caption{Plot of $R$ vs $\theta$. We have chosen $\bar{H}_{ini} = -1$ for
		simplification purposes, which implies that $t$ is in units of $1/\bar{H}_{ini}$. On
		the left side, we see the case for a large initial density contrast while the right
		side indicates a small initial density contrast. The blue and the orange lines
		indicate the positive and negative initial conditions for $\theta_{ini}$
		respectively. The blue line represents the right choice and a contracting model
		while the orange line indicates an expansion model. We chose different values of
		$\delta_{ini}$ for both graphs as the difference between both choices becomes more
		evident for a smaller initial density contrast. The data was computed with Wolfram
		Mathematica~\cite{Mathematica} and the graphs were generated in Python.}
	\label{theta_evol}
\end{figure}

In App.~\ref{appc}, we find the solutions for Einstein's equations of a critical
collapse represented with a local metric that depends on the local function $R(t,r)$,
which led to the Lemaître-Tolman-Bondi (LTB) set of solutions
\begin{align}
	\label{ltb3a}
	R(\theta, r) & = \frac{r (1+\delta_{ini})}{\delta_{ini}}\sin^2\left(\frac{\theta}{2}\right),                                        \\
	\label{ltb4a}
	t(\theta,r ) & = t_1(r) + \frac{1 + \delta_{ini}}{2\bar{H}_{ini}\delta^{\frac{3}{2}}_{ini}}\left(\theta - \pi - \sin \theta \right)
	,\end{align}
such that we use the notation $\delta_{ini} = \delta(t_{ini})$ at the initial collapse
time. In this section, we wish to impose some constraints on those solutions to find the
exact point at which the perturbations will form a black hole. Consequently, this will
lead to a critical value for the density contrast. First of all, let us analyze the
parameter $\theta$. From Eq.~\eqref{sintheta},
\begin{align}
	\label{thetadef}
	\theta_{ini} & = \pm 2\arcsin(\sqrt{\frac{\delta_{ini}}{1+\delta_{ini}}})
	.\end{align}
To choose between the positive or negative value, we use our initial condition in
Eq.~\eqref{hinicond} and Eq.~\eqref{hdefltb} as a test, such that replacing
Eq.~\eqref{thetadef}
\begin{align}
	\left.\frac{\dot{R}}{R}\right|_{ini} & =  \left.\frac{\frac{\partial R}{\partial \theta}}{R \frac{\partial t}{\partial \theta}}\right|_{ini} = \pm \bar{H}_{ini}.
\end{align}
From the above, we note that the positive sign in Eq.~\eqref{thetadef} leads to the
right definition of our initial conditions. If one chooses the negative sign, and as
$\bar{H}_{ini}$ is negative in the contracting phase, we would end up with an initially
expending patch. This feature can be seen in Fig.~\ref{theta_evol}, where
Eqs.~\eqref{ltb3a} and \eqref{ltb4a} were solved analytically with Wolfram
Mathematica~\cite{Mathematica}. The blue lines represent the positive choice that leads
to the LTB collapse model in a contracting universe, while the orange lines describe an
LTB collapse model with an initially expanding patch. On the left side of the figure,
for a larger value of $\delta_{ini}$, we see that both solutions are similar as the
orange plot has an insignificant initial expansion. For a smaller value of
$\delta_{ini}$ on the right, the model in orange first goes on an expansion phase
followed by the collapse, while the blue plot is already in a contracting phase and
collapses much earlier.

Let us now analyze the PBH formation. The collapsed object will form a singularity when
$\frac{\partial R}{\partial r} = {R}^\prime = 0$~\cite{Dey2023}. From Eq.~\eqref{ltb3a}
we see that
\begin{align}
	{R}^\prime & = \frac{ (1+\delta_{ini})}{\delta_{ini}}\sin^2\left(\frac{\theta}{2}\right) = 0
\end{align}
if $\theta = 0$. Consequently, we want to analyze the necessary time for a black hole to
form, which takes place at $\theta = 0$. From Eq.~\eqref{ltb4a} we have that the final
time $t_f$, i.e., the time at formation is
\begin{align}
	t_f & = t(0,r ) =t_1(r) - \frac{\pi(1 + \delta_{ini})}{2\bar{H}_{ini}\delta^{\frac{3}{2}}(t_{ini})}
	.\end{align}
We want to compute the formation time $\Delta t = t_f - t_i$, such that $t_i =
	t(\theta_{ini}, r)$ with
\begin{align}
	\theta_{ini} & = 2\arcsin(\sqrt{\frac{\delta_{ini}}{1+\delta_{ini}}})
	.\end{align}
Hence, from Eq.~\eqref{ltb4a}, we have that
\begin{align}
	\label{criticallinear}
	\Delta t & = t_f - t_i \nonumber                                                                                                                                                                                 \\
	         & =\frac{(1 + \delta_{ini})}{2\bar{H}_{ini}\delta^{\frac{3}{2}}_{ini} } \left(  -\theta_{ini} +\sin{\theta_{ini}}\right) ,\nonumber                                                                     \\
	         & =\frac{(1 + \delta_{ini})}{2\bar{H}_{ini}\delta^{\frac{3}{2}}_{ini} } \left( -2\arcsin(\sqrt{\frac{\delta_{ini}}{1+\delta_{ini}}}) +\sin(2\arcsin(\sqrt{\frac{\delta_{ini}}{1+\delta_{ini}}}))\right)
	.\end{align}
Assuming that $\delta_{ini} \ll 1$, we can expand the above relation such that
\begin{align}
	\label{deltasmall}
	\Delta t & = -\frac{2}{3\bar{H}_{ini}}+\frac{2\delta_{ini}}{15\bar{H}_{ini}} + \mathcal{O}^2(\delta_{ini})
	.\end{align}
We now know how to compute the time interval with the right side of the above equation.
We shall now study the left side of Eq.~\eqref{deltasmall} as we want to find a
constraint on the final formation time $t_f$. Supposing that the PBH is formed way
before the bounce\footnote{We do not make this approximation in the numerical code. This
	approximation is only used to demonstrate our calculations.}, we can approximate
Eq.~\eqref{bohm_hubble} to
\begin{align}
	\label{hubblefarbounce}
	\Bar{H} & = \frac{2}{3t}
	.\end{align}
Plugging this result for $t_i$ on the left side of Eq.~\eqref{deltasmall} leads to
\begin{align}
	\label{tf}
	t_f & = \frac{2\delta_{ini}}{15\bar{H}_{ini}}
	.\end{align}
Let us rewrite the above expression in terms of the redshift function $x$ defined as
\begin{align}
	x(t) & \equiv  \frac{\bar{a}_0}{\bar{a}(t)}
	,\end{align}
such that $\bar{a}_0$ is the scale factor today and the Hubble function can be written as
\begin{align}
	\label{Hx}
	\bar{H} & = -\bar{H}_0\sqrt{\Omega_w} x^{3/2}
	,\end{align}
being $\Omega_w$ the dust density in the universe today and thus Eq.~\eqref{tf} becomes
\begin{align}
	\label{tfx}
	t_f & = -\frac{2 \delta_{ini}}{15\bar{H}_0 \sqrt{\Omega_w} x_{ini}^{3/2}}
	.\end{align}

We need to impose a limit on the formation time to find the critical values of delta.
For collapsed objects larger than the Hubble radius, their dynamics will be dominated by
the FLRW metric and not the LTB approximation. The supper-Hubble perturbations will be
frozen and thus there is no collapse for these scales. This sets the Hubble length as an
upper cut-off, i.e., their formation time must be at most the time for the perturbation
to achieve the Hubble radius, labeled as $t_H$.  The radius is related to the
perturbation's wavelength $\lambda$ and comoving wavenumber $k$ by~\cite{Quintin2016}
\begin{align}
	\label{radius}
	r       & =\frac{\lambda}{2}, \\
	\label{wavelength}
	\lambda & = \frac{2 \pi}{k}.
\end{align}
We want to analyze when a perturbation with wavenumber $k$ has the same size as the
Hubble radius, that is,
\begin{align}
	\label{hubblek}
	k_H & = \frac{1}{x}\abs{\Bar{H}}
	.\end{align}
From Eq.~\eqref{hubblek} and Eq.~\eqref{Hx}, we have that the redshift function
associated with this scale is given by
\begin{align}
	\label{xh}
	x_H & = \frac{k_a^2 }{\Omega_w}
	,\end{align}
where $k_a$ is the comoving dimensionless wave number and the subscript $H$ indicates
the Hubble scale. In terms of time, we can rewrite Eq.~\eqref{xh} with
Eq.~\eqref{hubblefarbounce} and Eq.~\eqref{Hx} such that
\begin{align}
	t_H & = -\frac{2\Omega_w}{3 \bar{H}_0 k_a^2}
	.\end{align}
Finally, we must ensure that
\begin{align}
	\label{deltacf}
	t_f \leq t_H ~\text{or}\nonumber \\
	\delta_{ini} \geq \frac{5\Omega_w^{\frac{3}{2}} x_{ini}^{\frac{3}{2}}}{k_a^3}
	,\end{align}
where we used Eq.~\eqref{tfx} and Eq.~\eqref{kadm}. From the above, we can see that the
critical threshold depends both on the scale and time and it can be set from the
saturation of the inequality as
\begin{align}
	\label{deltacfinal}
	\delta_c & =
	\frac{5 \Omega_w^{\frac{3}{2}} x_{ini}^{\frac{3}{2}}}{k_a^3}
	.\end{align}
Let us now evaluate two cases: one considering the dark matter as a pressureless fluid
and another considering really small but finite pressure.

\subsection{$\Bar{p} = w \bar{\rho}$}

\begin{figure}[tbp]
	\centering
	\includegraphics[width=.6\textwidth]{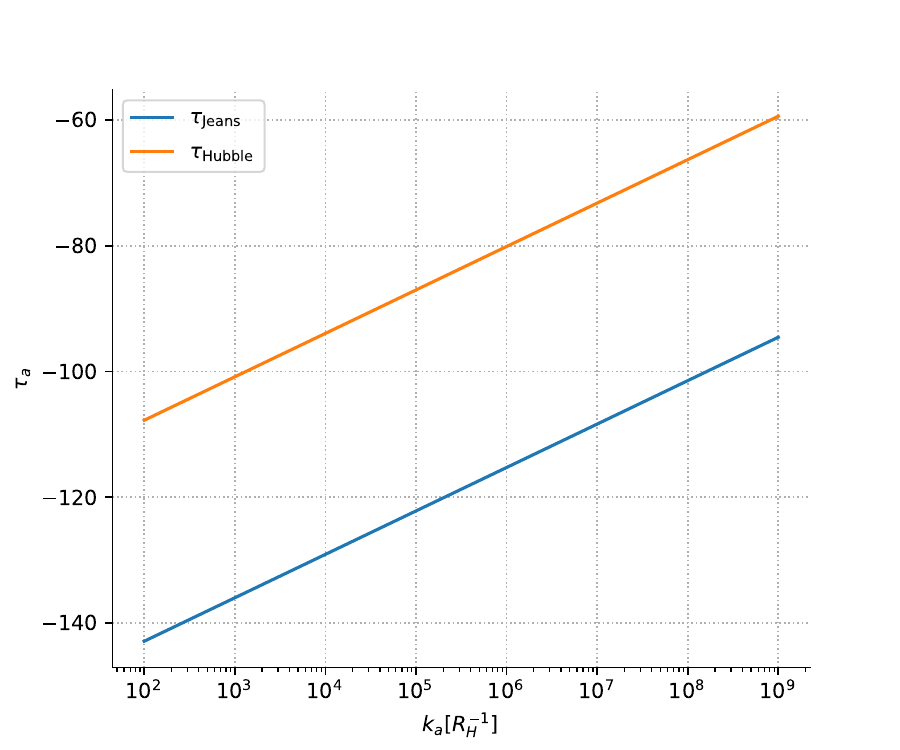}
	\caption{ Plot of the Jeans and Hubble time computed with the NumCosmo library for $w = 10^{-10}$ and $x_b = 10^{35}$. These times correspond to when the matter density perturbations reach either the Jeans or the Hubble length respectively for each mode $k$. The time parameter in the y-axis is given by Eq.~\eqref{tauparameter}.}
	\label{jeans_hubble_time}
\end{figure}

In the presence of pressure, it is well known that the Jean's length determines the
smaller scale sufficient for a black hole to be formed. The pressure forces oppose the
collapse and only for scales above this limit a black hole can be formed. Thus, we are
interested in the physical radius in Super-Jeans/Sub-Hubble scales, that is,
$r_j<r<r_H$. The Jeans comoving scale $k_j$ is given by~\cite{Quintin2016}
\begin{align}
	\label{jeansk}
	k_J & = \sqrt{\frac{3}{2}}\frac{1}{x} \frac{\abs{\Bar{H}}}{c_s}, \\
	.\end{align}
Thus, no structure smaller than the Jeans length can collapse and we have a lower
momenta cutoff $k<k_j$. Plugging this requirement plus our upper limit in
Eq.~\eqref{jeansk} leads to
\begin{align}
	\label{jeanslimit}
	k_H<k                                                                            & \leq k_j~\text{or}\nonumber                                     \\
	\left(\frac{1}{\sqrt{\Omega_w}}\right)^{\frac{3}{2}}>\frac{x^{\frac{2}{3}}}{k^3} & \geq \left(\frac{2c_s^2}{3\sqrt{\Omega_w}}\right)^{\frac{3}{2}}
\end{align}
where we have used Eq.~\eqref{hubblefarbounce} and Eq.~\eqref{Hx}. The above relation
indicates that for scales/times that do not satisfy the inequality, the perturbation
modes do not contribute to PBH formation. Hence we must evaluate the collapse starting
between the Jeans-Hubble time and always ending when the perturbations reach the Hubble
length. We can see in Fig.~\ref{jeans_hubble_time} the respective Jeans and Hubble time
for each different scale. Note that the interval between both times is always constant
for all modes, as it depends only on the dark matter equation of state $w$. Thus, for
smaller $w$, the gap between the times increases and the PBH formation is enhanced as
there is more time for them to be formed.

\newpage

Finally, we have the critical threshold in Eq.~\eqref{deltacfinal} with
Eq.~\eqref{jeanslimit} corresponding to the allowed scales. Hence, not all perturbations
will collapse into primordial black holes. Note that for smaller values of $c_s^2$,
there is more time between the Jeans and the Hubble scale, which allows for more
perturbations to collapse. Thus the abundance of PBH in this model is directly related
to the equation of state of dark matter, as we will see in the next section.

\begin{figure}[tbp]
	\centering
	\includegraphics[width=.6\textwidth]{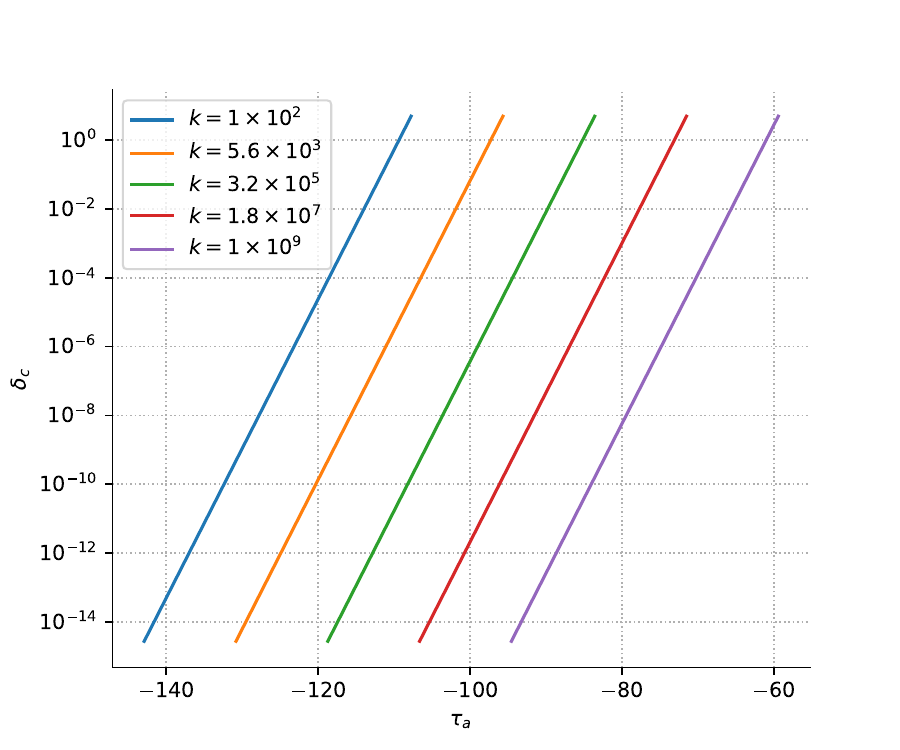}
	\caption{ Plot of the critical threshold vs time computed with the NumCosmo library
		for $w = 10^{-10}$ and $x_b = 10^{35}$. Each plot refers to a different scale $k$,
		while points on the same plot represents a different initial time for the collapse.
		The time interval for each mode starts at the Jeans scale, leading to the smallest
		threshold values, and ends for initial times when the perturbations reach the Hubble
		scale, which leads to higher threshold values. The time parameter in the x-axis is
		given by Eq.~\eqref{tauparameter}.}
	\label{deltacfig}
\end{figure}

In this context, let us now analyze Eq.~\eqref{deltacfinal} for the
super-Jeans/Sub-Hubble scales, depicted in Fig.~\ref{deltacfig}. Each plot begins for
$\delta_c$ being computed for $t_i = t_j$ and goes until $t_i = t_H$. The first case
corresponds to the collapse starting when the modes just achieved the Jeans length and
thus have the maximum amount of time to collapse until reaching the Hubble length. The
end of the time interval corresponds to a collapse taking place right before the
perturbation reaches the Hubble length, which leads to a maximum value of the critical
threshold since there is a minimum amount of time for the collapse of the perturbations.
We can see that all plots go from $\delta_c \approx 10^{-14}$ at Jean crossing time to
$\delta_c \approx 5.0$ at the Hubble scale. However, the closer we get to the Hubble
time as the initial time, the worse the approximation $\delta_c \ll 1$ gets and thus
this should be disregarded. If $t_i = t_H$, we have that $\delta_c \sim \infty$, since
there is no time for the collapse to happen. In Fig.~\ref{theta_evol} we have solved
Eq.~\eqref{deltacfinal} analytically for larger values of $\delta$. Nevertheless, such
values of density contrast will never be achieved during the contracting phase.

\newpage
\subsection{$\bar{p} = w \bar{\rho}, \quad w \ll 1$}

We want to analyze what would happen if the dark matter had such a small equation of
state that could be thought as a presureless fluid. In this case, Eq.~\eqref{deltacf}
tells us that, for every scale $k$, we can always find a sufficient time in the past
such that $x_{ini} \ll 1$, leading to the collapse of all density perturbations. In
other words, if the universe is old enough, all perturbations will eventually collapse
into a black hole until the universe reaches the bounce. If there was no bounce, the
universe would completely collapse into black holes and our model would be highly
unstable.

It is indeed expected that a fluid with no pressure leads to a total collapse since no
forces are opposing the collapse and dust only interacts via gravity. Thus all dark
matter that we see today would indeed be contained in primordial black holes. However,
this hypothesis was already refuted by some works by constraining the PBH abundance in
dark matter based on observational effects (see Refs.~\cite{Villanueva2021, Carr2021}).
Hence we must consider a dust fluid with a non-vanishing equation of state.

\subsection{Filtered Variance}

Let us now compute the variance which will determine the distribution for PBH. Since
sub-Jeans and super-Hubble scales do not contribute to the PBH formation that we are
interested in, we want to compute the integral in  Eq.~\eqref{eqsigmar} between $k_H$
and $k_j$. To do so, we first need to evaluate $P_{\tilde{\delta}}(k)$ using
Eqs.~\eqref{deltarho}-\eqref{phiz} and the field expansions in Eqs.~\eqref{vexp}
and~\eqref{pvexp}. Explicitly,
\begin{align}
	\label{psiq}
	\hat{\Psi}(\eta,\textbf{x}) & =\int\frac{\mathrm{d}^3\textbf{k}}{(2\pi)^{\frac{3}{2}}}
	\frac{\bar{H}}{2\Bar{a}k^2}\left[\Pi_{\zeta_{k}} e^{i\textbf{k}\textbf{x}} a_{k} +
	\Pi_{\zeta_{k}}^{ *}e^{-i\textbf{k}\textbf{x}} a_{k}^{\dagger}\right]
	,\end{align}
\begin{align}
	\bar{D}^2\hat{\Psi}(\eta,\textbf{x}) & =
	-\int\frac{\mathrm{d}^3\textbf{k}}{(2\pi)^{\frac{3}{2}}}
	\frac{\bar{H}}{2a^3 }\left[\Pi_{\zeta_{k}} e^{i\textbf{k}\textbf{x}} a_{k} +
	\Pi_{\zeta_{k}}^{*}e^{-i\textbf{k}\textbf{x}} a_{k}^{\dagger}\right]
\end{align}
and
\begin{align}
	\hat{\tilde{\delta\rho}}(\textbf{x}) & =
	\int_{-\infty}^{\infty}\frac{\mathrm{~d}^3\textbf{k}}{(2\pi)^{\frac{3}{2}}}
	\left\{ \left(\frac{\bar{H}}{\kappa \bar{a}^3}\right)\Pi_{\zeta_{k}} e^{i\textbf{k}\textbf{x}} a_{k} +\right. \nonumber                                     \\
	                                     & \left. \left(\frac{\bar{H}}{\kappa a^3}\right)\Pi_{\zeta_{k}}^{*} e^{-i\textbf{k}\textbf{x}} a_{k}^{\dagger}\right\}
	.\end{align}
Thus, the two-point function and the variance are given by
\begin{align}
	\label{2point}
	\left<\hat{\Tilde{\delta}}(\textbf{x})\hat{\Tilde{\delta}}(\textbf{y})\right> & =
	\frac{1}{2\pi^2} \int_0^\infty \frac{\mathrm{~d}k}{k}~\left[P_{\tilde{\delta}}(k)
		\frac{\sin{kR}}{kR} \right]
\end{align}
and
\begin{align}
	P_{\tilde{\delta}}(k) & = \abs{\Pi_{\zeta_{k}}}^2\left( \frac{1}{9\bar{a}^6 \Bar{H}^2(1+w)^2}\right),
\end{align}
where we used the flat Friedmann equation
\begin{align}
	\bar{H}^2 = \frac{\kappa \Bar{\rho}}{3}.
\end{align}
We are now able to compute the variance modes numerically using the above relation
together with Eq.~\eqref{physicalrho} in the fluid gauge and our numerical code.

\begin{figure}[tbp]
	\centering
	\includegraphics[width=.6\textwidth]{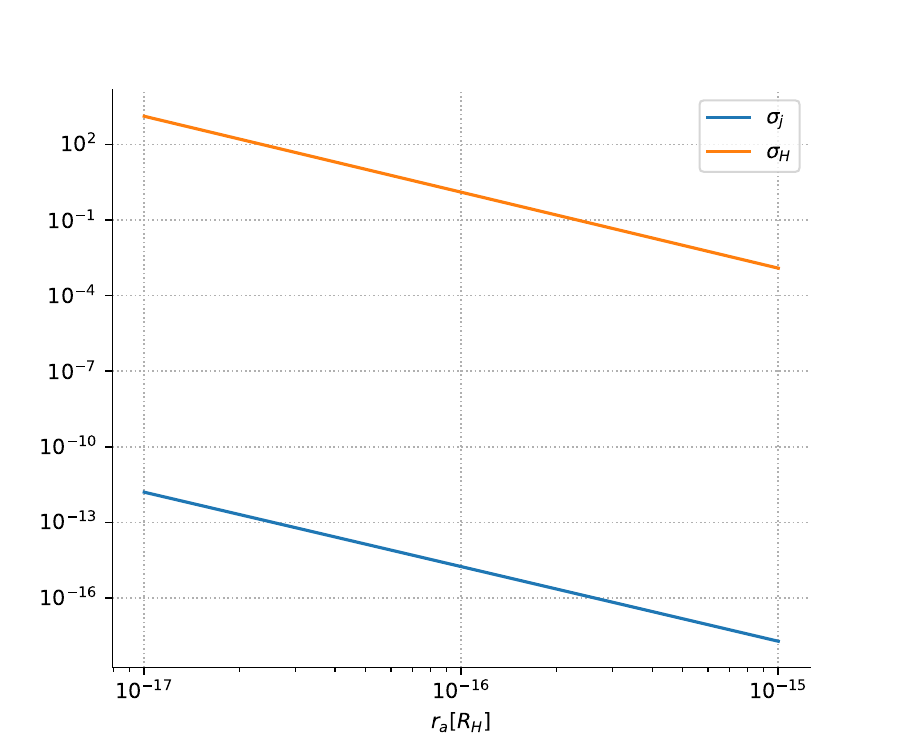}
	\caption{ Plot of the filtered variance computed with the NumCosmo library for $w =
			10^{-10}$ and $x_b = 10^{35}$. Each point corresponds to the variance computed when
		the perturbations become either the Jeans (blue plot) or the Hubble (orange plot)
		size for each scale.}
	\label{sigmas}
	.\end{figure}
\newpage

The values for the filtered variance computed with Eq.~\eqref{eqsigmar} from $k_H$ to
$k_j$ are displayed in Fig.~\ref{sigmas}. We have plotted the values of $\sigma_r$ for
two different times versus radius. The sub-indexes $H$ and $j$ indicate that we are
computing the variance at a time when the perturbations are the size of the Jeans length
or the Hubble horizon respectively. We can see a linear relation between the filtered
variance and the scale radius, such that the former decays with the latter. Larger
values of variance indicate a broader Gaussian distribution, which enhances the PBH
formation. Thus, from this figure, we see that the formation of PBHs on smaller scales
is prioritized rather than on larger scales. We now shall compute the PBH abundance in
this model.

\subsection{Mass Function}
\label{psf}

To analyze the abundance of collapsed objects, we use the Press-Schechter (PS)
formalism, first proposed in~\cite{press1974}. This ansatz presumes that the objects are
formed through a nonlinear collapse and have the property of being universal regarding
different cosmological models. Other works have proposed specific mass functions for the
primordial black hole production for specific mass ranges, for example~\cite{Cai2023,
	Bi2024}. However, in this work, we restrain ourselves to the universal PS formalism to
analyze all possible scales. Following this approach, the density of collapsed objects
per mass scale is given by the mass function
\begin{align}
	\label{ps}
	\frac{dn(z, M)}{dM} & = -\frac{\bar{\rho}_m(z)}{M}\frac{d\beta(z,M)}{dM}
	,\end{align}
where $\frac{dn(M,z)}{dM}$ is the mass function, $z$ is the object's redshift, $M$ is
the object's mass, and $\beta$ is the fraction of collapsed objects in the mass range of
$M+dM$.

Assuming that the objects will collapse when $\delta  > \delta_c$, we can measure the
mass fraction inside spheres of radius $r$ for a time $t$ that is constituted of
collapsed objects as
\begin{align}
	\label{betaf}
	\beta(t, r) \equiv \frac{\rho_{PBH}}{\rho} & =\int_{\delta_c}^{\infty}\mathrm{d}\delta~P(\delta_r) \nonumber \\
	                                           & = \text{erfc}\left(\frac{\delta_c}{\sqrt{2}\sigma_r(t)}\right)
	,\end{align}
where $erfc$ is the complementary error function. However, $\beta$ is a probability
density of how likely are perturbations to collapse after a time $t$. For instance, if
we compute it at the time when the perturbations reach the Hubble size, $\delta_c =
	\infty$, this integral will vanish, denoting that there are no more PBHs to be formed
after this time. In our context, we are interested in analyzing how many PBHs were
formed in the whole contracting phase so we can compute its abundance today. This
function shall be given by
\begin{align}
	\label{densityf}
	F(t, r)_{PBH} & \equiv \max(\beta(t_i,r)),~\text{for}~t_i \leq t
	.\end{align}
The maximum of the beta function indicates for which time there is a higher probability
of PBHs to be formed, which will be given by the Jeans time $t_{j}$. Thus $F(t, r)$ with
$t_i = t_j$ denotes the actual mass fraction of PBHs formed during the entire bouncing
phase. Hence, as an approximation, we will compute the mass fraction for the maximum
probability of forming PBHs, i.e., always beginning at $t_j$, to find the abundance of
PBHs today.

\begin{figure}[tbp]
	\centering
	\includegraphics[width=.6\textwidth]{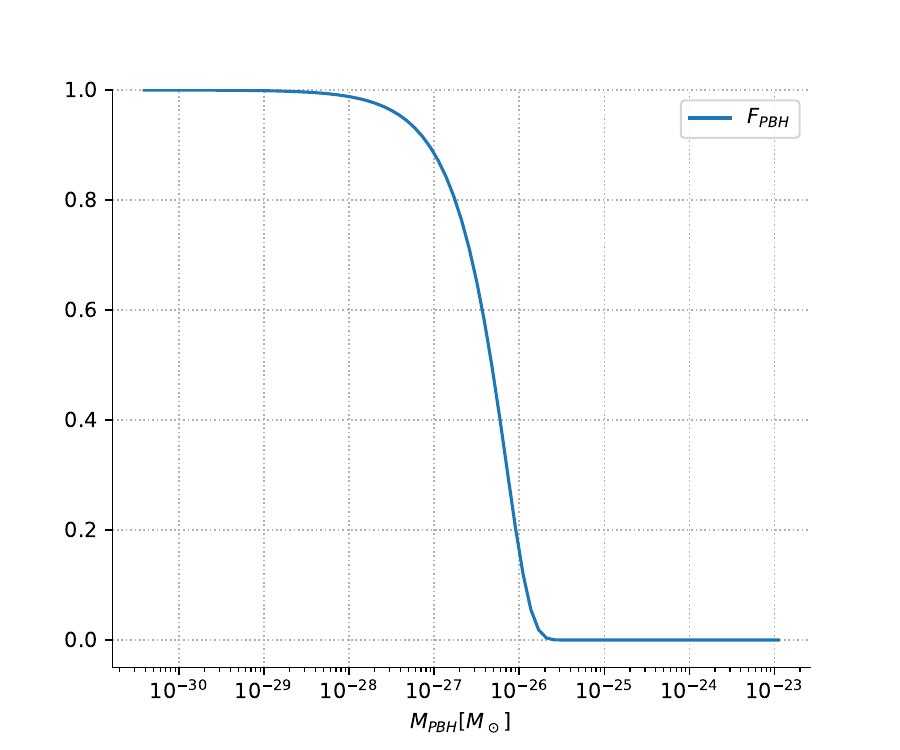}
	\caption{Plot of the mass fraction of PBHs forming during bounce versus their
		formation mass in solar masses for $w = 10^{-10}$ and $x_b = 10^{35}$. The
		computations were done using the NumCosmo library.}
	\label{fvsm}
\end{figure}

Another crucial quantity is the abundance of collapsed objects in the universe
$\Omega_{PBH}$, defined in Ref.~\cite{Carr2016} as the ratio between the density of PBHs
and the background universe's density today. This parameter will serve as a guide to
compare our theoretical predictions with observational data~\cite{Villanueva2021}. Since
both the background and the PBHs are formed by dust, both grow with the same power of
the scale factor and thus the PBH abundance today will be given by
\begin{align}
	\Omega_{PBH} & = F(t, r) \Omega_{DM} = \text{erfc}\left(\frac{\delta_c}{\sqrt{2}\sigma_r(t_{j})}\right) \Omega_{DM}
	.\end{align}
In Fig.~\ref{fvsm} we can see the values of the density function in Eq.~\eqref{densityf}
for different mass scales. Keep in mind that everything is being computed for the
collapse starting at the Jeans time for every scale.

\begin{figure}[tbp]
	\centering
	\includegraphics[width=.6\textwidth]{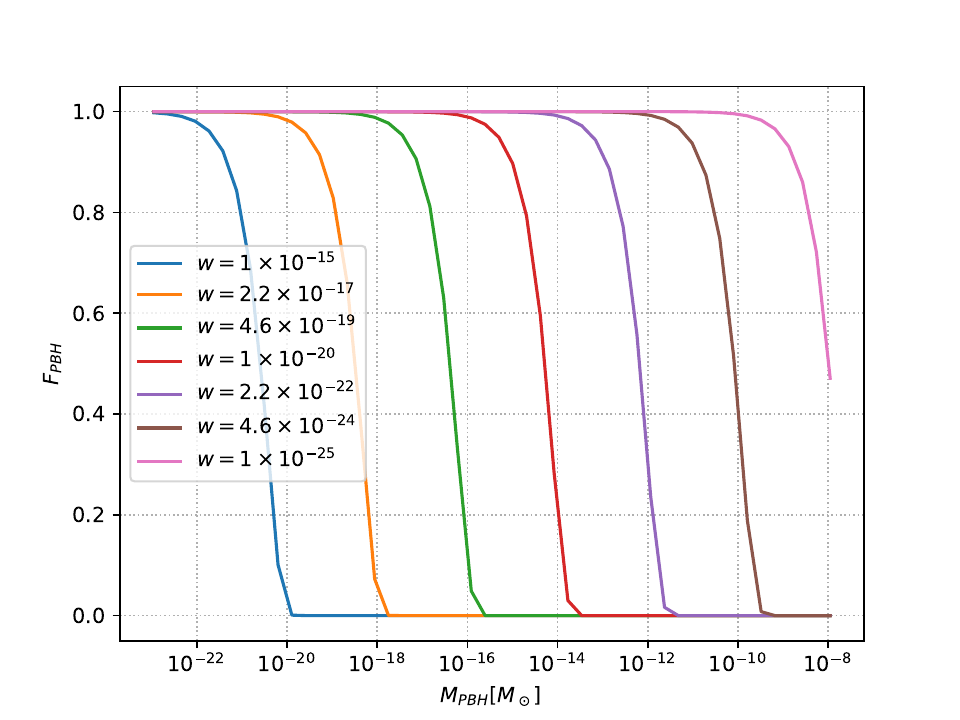}
	\caption{ Plot of the mass fraction of PBHs forming during bounce versus their
		formation mass in solar masses for different value of $w$ and $x_b = 10^{35}$. The
		computations were done using the NumCosmo library.}
	\label{fvsw}
\end{figure}

The complementary error function in Eq.~\eqref{betaf} has higher values when the
argument is closer to one, i.e., when the filtered variance and the critical density
contrast are in the same scale. We can see in Fig.~\ref{fvsm} that this is only true for
small mass scales, where the density function is equal to one. However, at these mass
scales, every formed PBH black hole would eventually evaporate. From
Ref.~\cite{Villanueva2021}, the evaporation constraint determines that only black holes
with masses $M>10^{-18}M_{\odot}$ would have not completely evaporated today. Thus, for
larger scales, the density fraction equals zero, and there is no relevant PBH formation
in this model. Nonetheless, since we are dealing only with dust, which has a small and
not completely determined equation of state, let us analyze the same function for
different values of $w$, depicted in Fig.~\ref{fvsw}.

For smaller values for the equation of state of cold dark matter, we can see an
enhancement in the density function.  This case reflects our previous discussion of a
pressureless fluid. As we go further into the smaller values, almost every scale will
collapse into a black hole and thus the density function equals one.


\section{Discussion and Conclusions}
\label{sec:discussion}

In this study, we investigated the formation of PBHs in a flat quantum bouncing model
containing only dark matter with a small equation of state parameter $w$. Despite the
spectra from this model exhibiting a slight blue tilt, which deviates from observations
of the Cosmic Microwave Background (CMB), this discrepancy should not preclude the
model, as the inclusion of radiation is expected to induce a different spectrum that may
be compatible with observations \cite{Vitenti2012}. The dark matter-only model serves as
an initial attempt to quantify PBH formation during the contracting phase.

Constraints for PBH formation were established, requiring their lengths to fall between
the Jeans and Hubble scales. These upper and lower bounds enabled the computation of the
critical density contrast, detailed in Sec.~\ref{critical_delta} and App.~\ref{appc}. We
obtained a time/scale-dependent critical threshold different from the usual constant due
to the contracting dynamics. This behavior was expected as there is not only one
characteristic scale for the collapse in the bounce model since the perturbations may
collapse for any scale between the Jeans and the Hubble scale, which affects the
threshold calculation. For the case of a pressureless field, perturbations for all
scales may collapse and the bounce/contracting phase duration would be the only
constraint for PBH formation.

\begin{figure}[tbp]
	\centering
	\includegraphics[width=1.0\linewidth]{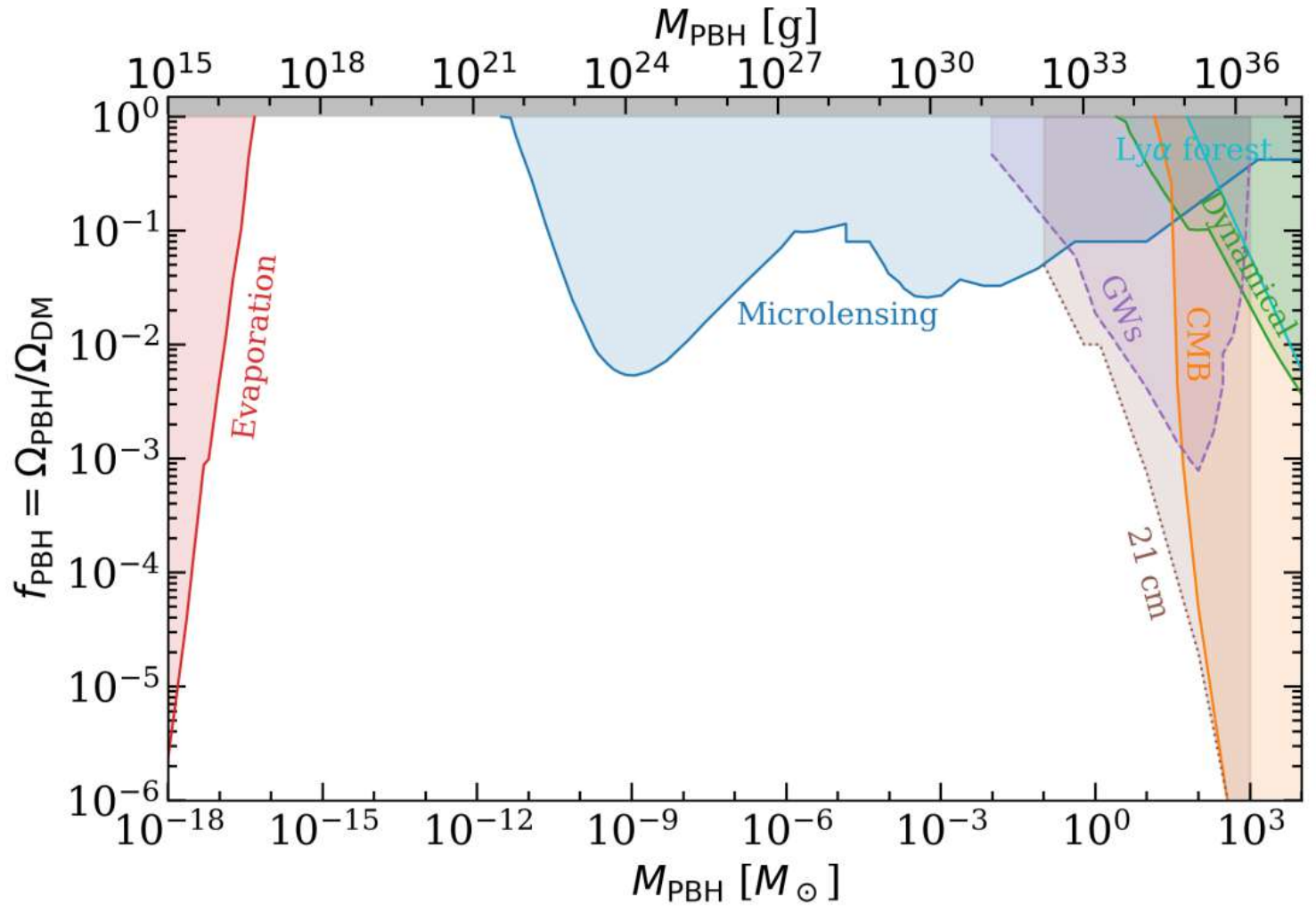}
	\caption{Plot of the density function equivalent to the one in Eq.~\eqref{densityf} versus the formation mass of PBHs. The colored areas are related to Constraints on the PBH fraction constituting dark matter today. Each color corresponds to a different probe considered to constrain PBH abundance. The uncolored area of the graph corresponds to the possible fraction values that agree with all probes. Figure from Ref.~\cite{Villanueva2021}.}
	\label{fig:pbhdm}
\end{figure}

For a small but non-zero pressure, we computed the mass fraction of primordial black
holes in the universe today, given by Eq.~\eqref{densityf}. In Fig.~\ref{fvsm}, for a
fixed value of $w$, we can see an enhancement of the density function for smaller mass
scales of PBHs. However, for $w = 10^{-10}$, the formed primordial black holes would
have such small masses at smaller scales that they would have evaporated completely
today. Figure~\ref{fig:pbhdm} was generated in Ref.~\cite{Villanueva2021} by imposing
observational constraints on the mass fraction of PBHs. They considered a range of
effects such as black hole evaporation, microlensing, gravitational wave measurements,
and others. The white regions of the graph correspond to accepted values of the mass
fraction and the colored represent physical observational effects that exclude the
possibility of PBHs constituting DM in that mass range. We can see that PBHs under the
mass range $M <  10^{-18}M_\odot$ are disregarded due to evaporation, which implies that
the formed PBHs depicted in Fig.~\ref{fvsm} would have completely evaporated today.

Still on Fig.~\ref{fvsm}, for larger scales, there is no significant formation as the
time interval for the collapse is not sufficient nor is the variance large enough. Also,
the variance decreases with scale, as seen in Fig.~\ref{sigmas}, while the critical
threshold remains constant as it is always computed for the Jeans time at the given
scale. Hence the complementary error function decreases since the argument grows and the
PBH fraction becomes insignificant.

We have seen in Eq.~\eqref{jeanslimit} that the time interval for which the
perturbations become super-Jeans/sub-Hubble is proportional to $c_s^3 =
	w^{\frac{3}{2}}$. Thus smaller values for the equation of state allow for more time for
the perturbations to collapse and therefore enhance the primordial black hole formation.
In Fig.~\ref{fvsw}, we see that only models featuring sufficiently small values for the
equation of state of dark matter ($w < 10^{-17}$) may lead to a non-vanishing mass
fraction of primordial black hole at relevant scales ($M >  10^{-18}M_\odot$). If we go
even further on small values of $w$, more scales start to collapse as we have $F\sim 1$
for all scales as we approach the pressureless fluid scenario.

Another approach for dark matter would be to consider a non-zero temperature. In
Ref~\cite{Armendariz2014} dark matter is treated as a non-relativistic gas and, even for
cases where $w\ll1$, the non-vanishing temperature would erase structure formation for
scales smaller than the corresponding free-streaming length, which once again would
contribute to decrease PBH formation in this model. If we treat DM as a relativistic
gas, its temperature grows with $\Bar{a}^{-2}$~\cite{Mukhanov1992}. Thus, its
temperature and pressure would diverge close to the bounce and prevent the collapse and
PBH formation. We conclude that the formation of PBHs in the flat-dust bounce in
observable scales is improbable as the analysis is robust against the formation on
larger scales, leading to an insignificant fraction of PBHs as DM today.

In this work, we have investigated the formation of primordial black holes within a
Friedmann background, incorporating scalar perturbations. An alternative approach would
be to explore PBH formation in a homogeneous yet anisotropic background. The presence of
anisotropy could give rise to PBHs with significant angular momentum, contrasting with
those formed in standard inflationary scenarios. A study of this possibility is left for
future work. The next phase of this research would be to apply the same methodology to a
universe populated with radiation and dark matter, which more closely resembles the
physical universe and the interaction between both fluids may affect PBH production. If
such a model proves reasonable, it would also be interesting to improve the computation
of the LTB model in App.~\ref{appc} for larger values of the equation of state. Despite
the Jeans scale being a good approximation, the introduction of a non-pressureless fluid
would also require the study of the pressure impact in our characteristic scales.
Furthermore, in a radiation-dominated quantum bouncing model, one expects a greater blue
tilt on the spectral index, which may affect the density contrast modes for smaller
scales. Said effect could lead to a larger formation of PBHs. This analysis is left for
future work.



\appendix

\section{Einstein Equations for Spherical Collapse}
\label{appc}

In this Appendix, we are interested in solving the Einstein equations for a metric
representing the spherical collapse. We intend to use these solutions to compute the
critical threshold in Sec.~\ref{critical_delta}. We adapted the calculations initially
done in Ref.~\cite{Gonccalves2000} and corrected by Ref.~\cite{Martin2020}. In these
references, a scalar field is used as the source for the energy-momentum tensor. We
shall now perform the same calculations for a pressureless barotropic fluid.

An inhomogeneous but spherically symmetric space can be represented by the
metric\footnote{This metric can also be written with a lapse function that shifts the
	time component. However, one can always redefine the coordinate to incorporate this
	quantity.}
\begin{align}
	\mathrm{d} s^2=-\mathrm{d}t^2+e^{-2 \Lambda(t, r)} \mathrm{d} r^2+R^2(t, r)\left(\mathrm{d} \theta^2+\sin ^2 \theta \mathrm{d} \varphi^2\right)
	,\end{align}
where $\Lambda(r,t)$ and $R(t,r)$ are functions of the local coordinates to be defined
and we are considering spherical coordinates. Since none of these functions do not
depend on the angular variables, this metric may represent any type of spherical
collapse. To find the local functions that describe our problem, we need to solve
Einstein's equations for an energy-momentum tensor that corresponds to our symmetries.
We shall represent $\partial_r \equiv {}^\prime$.

Following \cite{Martin2020}, the Einstein's tensor components are
\begin{align}
	G_{t t}           & =\frac{1}{R^2}\left[1+\dot{R}^2-2 \dot{\Lambda} \dot{R} R-R e^{2 \Lambda}\left(2 \Lambda^{\prime} R^{\prime}+2 R^{\prime \prime}+\frac{R^{\prime 2}}{R}\right)\right]                            \\
	G_{t r}           & =-\frac{2}{R}\left(\dot{R}^{\prime}+\dot{\Lambda} R^{\prime}\right)                                                                                                                              \\
	G_{r r}           & =\frac{1}{R^2}\left[R^{\prime 2}-e^{-2 \Lambda}\left(\dot{R}^2+2 R \ddot{R}+1\right)\right]                                                                                                      \\
	G_{\theta \theta} & =\sin ^{-2} \theta G_{\varphi \varphi}=R\left(\dot{R} \dot{\Lambda}+\Lambda^{\prime} R^{\prime} e^{2 \Lambda}+R^{\prime \prime} e^{2 \Lambda}-\ddot{R}+\ddot{\Lambda} R-R \dot{\Lambda}^2\right)
	.\end{align}
We want to consider an isotropic and inhomogeneous barotropic fluid. Thus our
energy-momentum tensor will have the form of a perfect fluid
\begin{align}
	\label{perfectfr}
	T_{\mu\nu} & = (\rho(r,t)+ p(r,t) )u_\mu u_\nu + p(r,t) g _{\mu\nu}
\end{align}
where $u_{\mu} u^\mu = -1$ is the velocity vector of the fluid, which we assume to be
orthogonal to the spatial hypersurfaces. It is worth mentioning that in
Eq.~\eqref{perfectfr}, both the pressure and the energy density depend on the spatial
radius since we want to consider a physical metric modeled as a perturbation around the
background quantities. Also, the equation of state parameter of the fluid is
\begin{align}
	\label{eqst}
	w & = \frac{p}{\rho}
\end{align}
and we are considering $w \ll 1$ for cold dark matter. In this context, the
energy-momentum tensor components are
\begin{align}
	\label{emprojec}
	T_{u u}     & =\rho(t, r)                                         \\
	T^{r}{}_r   & =T^{\theta}{}_\theta=T^{\varphi}{}_\varphi =p(t, r) \\
	T^{r}{}_{u} & = 0,                                                \\
	T_{r u}     & = 0
	,\end{align}
where the indexes $u$ and $r$ indicate both the fluid's velocity direction and the
radial direction respectively. Let us now move to the Einstein's equations (EE).

We can redefine our variables as
\begin{align}
	\label{kmdef}
	k(t, r)=1-R^{\prime 2} e^{2 \Lambda}, \quad m(t, r)=\frac{R}{2}\left(\dot{R}^2+k\right)
	,\end{align}
such that the EE will be given by
\begin{align}
	 & k^{\prime} = \kappa RR^{\prime}\left(T_{uu}+T^{r}{}_{r}\right)+2R^{\prime}\left(\ddot{R}+\dot{\Lambda}\dot{R}\right), \\
	 & \dot{k} = \kappa RR^{\prime}T^{r}{}_{u},                                                                              \\
	 & m^{\prime} =\frac{\kappa}{2} R^{2}R^{\prime}T_{uu}-\frac{\kappa}{2} R^{2}\dot{R}T_{ru},                               \\
	 & \dot{m} =\frac{\kappa}{2} R^{2}R^{\prime}T^{r}{}_{u}-\frac{\kappa}{2}\dot{R}R^{2}T^{r}{}_{r}
	.\end{align}
Additionally, the energy conservation of $T^{\mu\nu}$ gives the constraint
\begin{align}
	\label{eqenergy}
	 & \nabla_{\mu}T^{\mu\nu}=0,~\text{or}\nonumber                                                 \\
	 & \nabla_\mu(\rho u^\mu u^\nu+p\mathrm{~h}^{\mu\nu})=\left(\dot{\rho}+3H(\rho+p)\right)u^\nu=0
	,\end{align}
where $\mathrm{~h}^{\mu\nu} = g^{\mu\nu} + u^\mu u^\nu $.
Thus, our system is given by
\begin{align}
	\label{metricgen}
	\mathrm{d} s^2=-\mathrm{d} t^2+\frac{(R^\prime)^{2}}{1 - k} \mathrm{d} r^2+R^2(t, r)\left(\mathrm{d} \theta^2+\sin ^2 \theta \mathrm{d} \varphi^2\right)
\end{align}
with
\begin{align}
	\label{kprime}
	k^{\prime} & =\kappa RR^{\prime}\left(\rho+p\right)+2R^{\prime}\left(\ddot{R}+\dot{\Lambda}\dot{R}\right), \\
	\label{kdot}
	\dot{k}    & = 0,                                                                                          \\
	m^{\prime} & =\frac{\kappa}{2} R^{2}R^{\prime}\rho                                                         \\
	\label{mdot}
	\dot{m}    & =-\frac{\kappa}{2}\dot{R}R^{2}p = -\frac{\kappa}{2}\dot{R}R^{2} w\rho = 0.
\end{align}
plus Eqs.~\eqref{eqst} and~\eqref{eqenergy}. One can replace Eqs.~\eqref{mdot} and
\eqref{kdot} into \eqref{kprime} using the definitions in \eqref{kmdef} yielding the
final equations
\begin{align}
	\label{mprime}
	m^{\prime}  & =\frac{\kappa}{2} R^{2}R^{\prime}\rho \\
	\label{rdot}
	(\dot{R})^2 & = \frac{2m}{R}-k
	.\end{align}

We can recognize in Eqs.~\eqref{mprime} and \eqref{rdot} the Lemaitre-Tolman-Bondi (LTB)
class of solutions~\cite{Lemaitre1933}. One may check that the metric in
Eq.~\eqref{metricgen} describes the Schwarshild metric if $m$ is constant, the
Einstein-de Sitter universe if $R = a(t)r$ and $k=0$ and the closed Friedmann universe
if $R= a(t)r$ and $k = r^2$. We want to find analytical solutions for the system without
considering a specific type of these local functions. Luckily, the LTB class of
solutions is one of the few cases where there is an analytical parametric solution to
Einstein's field equations, given by
\begin{align}
	\label{ltb1}
	R(\theta, r) & = \frac{2m}{k}\sin^2\left(\frac{\theta}{2}\right)                          \\
	\label{ltb2}
	t(\theta, r) & = t_1(r) + \frac{m}{k^{\frac{3}{2}}}\left(\theta - \pi - \sin\theta\right)
	.\end{align}
In the above, $\theta$ is a parameter in the range $[-2\pi, 0]$ and $t_1$ is an
integration constant\footnote{Our definition of the parameter $\theta$ differs from
	Ref.~\cite{Martin2020} by $-\pi$ for simplification purposes.}. We still have to find
$m$ and $k$, which will require initial conditions. Let us first analyze $m$.

From the right side of Eq.~\eqref{mprime}, we see that $m$ is related to the mass
density of a spherical volume. To analyze this quantity, we start by rewriting the
energy density from the spherical collapse metric as
\begin{align}
	\rho(t, r) & = \bar{\rho}(t)(1 + \delta(t,r))
	,\end{align}
such that $\delta$ is the density contrast $\delta = \frac{\rho(t,r) -
		\bar{\rho}(t)}{\bar{\rho}(t)}$ and $\bar{\rho}$ the background energy density,
 note that this is the same density contrast as in
		Eq.~\eqref{densitycon} but using the approximation $w \ll 1$ for cold dark
		matter. We consider that there is a small overdensity in our spherical system
when compared to the background density. We want to study a top-hat spherical
collapse so that outside the over-dense spherical shell characterized by a
critical radius $r_c$, the energy density is approximately the background
density, i.e., ($r > r_c$)$\rightarrow$ $\rho \approx \bar{\rho}$. We can add
this information to the density contrast by making the redefinition
\begin{align}
	\delta(t, r) & \rightarrow \delta(t, r) \Theta(r-r_c)
\end{align}
where $\Theta$ is the Heaviside function. In this context, Eq.~\eqref{mprime} becomes
\begin{align}
	m(r) & = \int_0^r dr_1~ 4\pi GR^2R^\prime \left(\bar{\rho} + \delta \Theta(r-r_c)\right).
\end{align}

We still have freedom in our metric to define the local function $R$, which can be fixed
with initial conditions. The simplest choice would be that at initial collapse time
$t_{ini}$, $R(t_{ini},r) = r$, which implies that the 3-dimensional spherical shell is
initially at rest\footnote{Note that any other initial condition is valid and sufficient
	to completely define $m$.}. This leads to
\begin{align}
	\label{mr1}
	m(r) & = \frac{4\pi r^3G \bar{\rho}(t_{ini})}{3}\left(1 + \frac{3}{r^3}\int_0^r dr_1 (r_1)^2\delta(t_{ini}, r_1)\Theta(r-r_c)\right).
\end{align}
The integral represents the expected value of the density contrast inside a spherical
region with radius $r$ and $m$ gives the total mass inside this region. To evaluate it,
we need to assume that at the initial time, our spherical metric possesses the same
symmetries as the background and thus it is homogeneous. This means that the density
contrast has a uniform distribution and does not depend on position at
$t_{ini}$\footnote{This is only true at $t_{ini}$. At later times, the perturbation
	evolves and depends on the radial position.}. Consequently,
\begin{align}
	\label{deltaini}
	\delta(t_{ini}, r) & = \delta(t_{ini}) \equiv  \delta_{ini}.
\end{align}
We are representing any variable $V$ at initial time $t_{ini}$ as $V(t_{ini})\equiv
	V_{ini}$ to simplify the notation. Plugging Eq.~\eqref{deltaini} in \eqref{mr1} leads to
\begin{align}
	\label{pbhmassmetric}
	m (r) & =  \left\{\begin{array}{ll}
		                  M\left(\frac{r^3}{r_c^3}\right)                                       & r \leq r_c \\
		                  M+\frac{M}{1+\delta_{ini}}\left(\frac{r^3}{r_{\mathrm{c}}^3}-1\right) & r > r_c
	                  \end{array}\right.,
\end{align}
such that we rewrote the terms as a function of
\begin{align}
	M & =m(r_c)=\frac{4\pi G \bar{\rho}_{ini}r^3_c}{3}\left(1 + \delta_{ini}\right)
	.\end{align}

We may now compute $k$ using the second relation in Eq.~\eqref{kmdef}, which requires us
to evaluate $\dot{R}(t, r)$. To do so, we define the inhomogenous Hubble function
\begin{align}
	\label{hdefltb}
	H & \equiv \frac{\dot{R}(t, r)}{R(t, r)}
	.\end{align}
Since we have the vanishing relations in Eqs.~\eqref{mdot} and \eqref{kdot}, $R$ is the
only local degree of freedom that evolves in time and provides the dynamics of the
universe according to Eq.~\eqref{ltb1}. Hence it is natural that we use this quantity to
define the Hubble function.  Assuming that the local metric is homogeneous at $t_{ini}$,
\begin{align}
	\label{hinicond}
	H^2(t_{ini}, r) & = \bar{H}^2(t_{ini}) = \frac{ \kappa \bar{\rho}}{3} = \frac{2 M}{1 +  \delta_{ini}} \frac{1}{r_c^3}
	.\end{align}
Thus, we can use this relation to compute $k$ for $t = t_{ini}$ since this variable does
not evolve in time, which leads to
\begin{align}
	k(r) & = \frac{2m}{r} - r^2H^2_{ini}= \left\{\begin{array}{ll}
		                                             2M\frac{\delta_{ini}}{1+\delta_{ini}}\frac{r^2}{r_\mathrm{c}^3} & r \leq r_c \\
		                                             2M\frac{\delta_\mathrm{ini}}{1+\delta_{ini}}\frac{1}{r}         & r > r_c
	                                             \end{array}\right.
\end{align}

Now that we have properly computed our local quantities, we can rewrite our solutions in
Eq.~\eqref{ltb1} and~\eqref{ltb2} for $r\leq r_c$ as
\begin{align}
	\label{ltb3}
	R(\theta, r)        & = \frac{r (1+\delta_{ini})}{\delta_{ini}}\sin^2\left(\frac{\theta}{2}\right),                             \\
	\label{ltb4}
	\Delta t(\theta,r ) & = \frac{1 + \delta_{ini}}{2H_{ini}\delta^{\frac{3}{2}}(t_{ini})}\left(\theta - \pi - \sin \theta \right),
\end{align}
such that $\Delta t = t_i - t_j$ for arbitrary times and $\theta_{ini}$ is obtained from
Eq.~\eqref{ltb1} at $t_{ini}$, i.e,
\begin{align}
	\label{sintheta}
	\sin^2\left(\frac{\theta_{ini}}{2}\right) & = \frac{\delta_{ini}}{1 + \delta_{ini}}
	.\end{align}
We are now able to use these solutions to compute the critical value for the density
contrast.

\section{Fluid's Gauge}
\label{appgauge}

In the last appendix, we computed the solution for the Einstein equations for an
inhomogeneous dust-perfect fluid with a spherically symmetric local metric. To use this
solution and compare it with our perturbations defined in Sec.~\ref{linearpert}, we must
ensure that measurements in the local metric are in the same Gauge as our perturbations.
Since we projected the energy-momentum tensor in the direction of the velocity of the
fluid in Eq.~\eqref{emprojec}, we consider the Gauge where the fluid is at rest and we
shall now see how to define our perturbation theory in this Gauge choice.

Essentially, following \cite{Vitenti2014covariant}, we want to make sure that the local tensor in \eqref{perfectfr} can be seen as the perturbed energy-momentum tensor $\delta T_{\mu\nu}$. Explicitly, we want
\begin{align}
	T_{\mu \nu}\propto \delta T_{\mu \nu}=(\delta \rho-2 \phi) n_\mu n_\nu+2(\rho+p) n_{(\mu} \Bar{D}_{\nu)}\mathcal{V}+ \left(\delta p \right)\mathrm{~h}_{\mu \nu}+2 p\left(n_{(\mu} \Bar{D}_{\nu)}\mathcal{B}+\Bar{D}_\mu \Bar{D}_\nu \mathcal{E}\right)
	,\end{align}
where we consider no anisotropic pressure. Thus,
the fluid's Gauge is obtained by setting
\begin{align}
	\label{psigauge}
	\mathcal{V}             & = \mathcal{E}= 0 ~~\text{and} \\
	\mathcal{B}\rvert_{t_1} & = 0
	.\end{align}
With these Gauge choices,
\begin{align}
	\label{tensorconect}
	\delta T_{\mu \nu}=(\delta \rho-2 \phi) n_\mu n_\nu+ \left(\delta p \right)\mathrm{~h}_{\mu \nu}+2 p\left(\Bar{D}_\mu \Bar{D}_\nu \mathcal{E}\right)
	.\end{align}

The first choice for $\mathcal{V}$ in Eq.~\eqref{psigauge} assures that the perturbed
fluid is at rest with the background universe. The other two conditions guarantee that,
at least for an initial time $t_1$, there are no off-diagonal terms and we have an
isotropic perturbed fluid. For different times, it is not possible to set a Gauge such
that $\mathcal{E}=0$ for $t \neq t_1$~\cite{vitenti2012large}. Nonetheless, we can argue
that this term is proportional to the pressure of the fluid which has an almost
vanishing value for cold dark matter and thus can be discarded. Also, since our main
goal is to make proper measurements of the density contrast that depends on
Eq.~\eqref{deltarhoinvariant}, this term is not relevant to our computations. Finally,
in the fluid's Gauge, the gauge-invariant density contrast may be interpreted as its
physical equivalent, that is,
\begin{align}
	\label{physicalrho}
	\Tilde{\delta \rho} & = \delta \rho + \mathcal{V}\dot{\Bar{\rho}} = \delta \rho\end{align}
where we used Eq.~\eqref{deltarhoinvariant}.

\acknowledgments

SDPV acknowledges the support of CNPq of Brazil under grant PQ-II 316734/2021-7. EJB
acknowledges the support of CAPES under the grant DS 88887.510837/2020-00. LFD
acknowledges the support of CAPES under the grant DS 88887.902808/2023-00. We thank
Nelson Pinto-Neto and Sheng-Feng Yan for their valuable discussions.

\bibliographystyle{unsrt}
\bibliography{g}

\begin{thebibliography}{10}

\bibitem{Zel1967}
Ya.~B. {Zel'dovich} and I.~D. {Novikov}.
\newblock {The Hypothesis of Cores Retarded during Expansion and the Hot Cosmological Model}.
\newblock {\em Soviet Astronomy}, 10:602, February 1967.

\bibitem{Hawking1971}
Stephen Hawking.
\newblock Gravitationally collapsed objects of very low mass.
\newblock {\em Monthly Notices of the Royal Astronomical Society}, 152(1):75--78, 1971.

\bibitem{Hawking1974}
Stephen~W Hawking.
\newblock Black hole explosions?
\newblock {\em Nature}, 248(5443):30--31, 1974.

\bibitem{Carr1974}
Bernard~J Carr and Stephen~W Hawking.
\newblock Black holes in the early universe.
\newblock {\em Monthly Notices of the Royal Astronomical Society}, 168(2):399--415, 1974.

\bibitem{Carr1975}
B.~J. Carr.
\newblock {The primordial black hole mass spectrum.}
\newblock {\em Astrophys. J.}, 201:1--19, October 1975.

\bibitem{Hee1996}
Il~Hee~Kim and Chul~H Lee.
\newblock Constraints on the spectral index from primordial black holes.
\newblock {\em Physical Review D}, 54(10):6001, 1996.

\bibitem{Ricotti2008}
Massimo Ricotti, Jeremiah~P Ostriker, and Katherine~J Mack.
\newblock Effect of primordial black holes on the cosmic microwave background and cosmological parameter estimates.
\newblock {\em The Astrophysical Journal}, 680(2):829, 2008.

\bibitem{Dom2021}
Guillem Dom{\`{e}}nech, Volodymyr Takhistov, and Misao Sasaki.
\newblock Exploring evaporating primordial black holes with gravitational waves.
\newblock {\em Physics Letters B}, 823:136722, dec 2021.

\bibitem{Wang2022}
Sai Wang and Zhi-Chao Zhao.
\newblock {GW}200105 and {GW}200115 are compatible with a scenario of primordial black hole binary coalescences.
\newblock {\em The European Physical Journal C}, 82(1), jan 2022.

\bibitem{1975Natur}
G.~F. {Chapline}.
\newblock {Cosmological effects of primordial black holes}.
\newblock {\em Nature}, 253(5489):251--252, January 1975.

\bibitem{Cyburt2003}
Richard~H Cyburt, Brian~D Fields, and Keith~A Olive.
\newblock Primordial nucleosynthesis in light of wmap.
\newblock {\em Physics Letters B}, 567(3-4):227--234, 2003.

\bibitem{Abbott2016}
{The LIGO Scientific Collaboration, the Virgo Collaboration}.
\newblock {Observation of gravitational waves from a binary black hole merger}.
\newblock {\em Phys. Rev. Lett.}, 116(6):1--16, 2016.

\bibitem{Abbott2019}
BP~Abbott, R~Abbott, TD~Abbott, S~Abraham, Fausto Acernese, K~Ackley, C~Adams, Rana~X Adhikari, VB~Adya, C~Affeldt, et~al.
\newblock Binary black hole population properties inferred from the first and second observing runs of advanced ligo and advanced virgo.
\newblock {\em The Astrophysical Journal Letters}, 882(2):L24, 2019.

\bibitem{Escriva2023}
Albert Escrivà, Florian Kuhnel, and Yuichiro Tada.
\newblock Primordial black holes, 2023.

\bibitem{Passaglia2022}
Samuel Passaglia and Misao Sasaki.
\newblock Primordial black holes from {CDM} isocurvature perturbations.
\newblock {\em Physical Review D}, 105(10), may 2022.

\bibitem{Khlopov1998}
M.~Yu. Khlopov, R.~V. Konoplich, S.~G. Rubin, and A.~S. Sakharov.
\newblock Formation of black holes in first order phase transitions, 1998.

\bibitem{Liu2022}
Jing Liu, Ligong Bian, Rong-Gen Cai, Zong-Kuan Guo, and Shao-Jiang Wang.
\newblock Primordial black hole production during first-order phase transitions.
\newblock {\em Physical Review D}, 105(2), jan 2022.

\bibitem{Niemeyer1998}
Jens~C Niemeyer and Karsten Jedamzik.
\newblock Near-critical gravitational collapse and the initial mass function of primordial black holes.
\newblock {\em Physical Review Letters}, 80(25):5481, 1998.

\bibitem{Starobinskii1979}
A.~A. Starobinskii.
\newblock Spectrum of relict gravitational radiation and the early state of the universe.
\newblock {\em ZhETF Pis ma Redaktsiiu}, 30:719--723, 1979.

\bibitem{Guth1981}
Alan~H. Guth.
\newblock The inflationary universe: A possible solution to the horizon and flatness problems.
\newblock {\em Phys. Rev. D}, 23:347--356, 1981.

\bibitem{Bardeen1983}
James~M Bardeen, Paul~J Steinhardt, and Michael~S Turner.
\newblock Spontaneous creation of almost scale-free density perturbations in an inflationary universe.
\newblock {\em Physical Review D}, 28(4):679, 1983.

\bibitem{Linde1982}
Andrei~D Linde.
\newblock A new inflationary universe scenario: a possible solution of the horizon, flatness, homogeneity, isotropy and primordial monopole problems.
\newblock {\em Physics Letters B}, 108(6):389--393, 1982.

\bibitem{Bullock1997}
James~S Bullock and Joel~R Primack.
\newblock Non-gaussian fluctuations and primordial black holes from inflation.
\newblock {\em Physical Review D}, 55(12):7423, 1997.

\bibitem{Yokoyama1998}
Jun’ichi Yokoyama.
\newblock Chaotic new inflation and formation of primordial black holes.
\newblock {\em Physical Review D}, 58(8):083510, 1998.

\bibitem{Josan2010}
Amandeep~S Josan and Anne~M Green.
\newblock Constraints from primordial black hole formation at the end of inflation.
\newblock {\em Physical Review D}, 82(4):047303, 2010.

\bibitem{Ballesteros2018}
Guillermo Ballesteros and Marco Taoso.
\newblock Primordial black hole dark matter from single field inflation.
\newblock {\em Physical Review D}, 97(2):023501, 2018.

\bibitem{Wang2024}
Xinpeng Wang, Ying-li Zhang, and Misao Sasaki.
\newblock Enhanced curvature perturbation and primordial black hole formation in two-stage inflation with a break.
\newblock {\em arXiv preprint arXiv:2404.02492}, 2024.

\bibitem{Carr2018}
Bernard Carr, Konstantinos Dimopoulos, Charlotte Owen, and Tommi Tenkanen.
\newblock Primordial black hole formation during slow reheating after inflation.
\newblock {\em Physical Review D}, 97(12):123535, 2018.

\bibitem{Martin2020}
J{\'e}r{\^o}me Martin, Theodoros Papanikolaou, and Vincent Vennin.
\newblock Primordial black holes from the preheating instability in single-field inflation.
\newblock {\em Journal of Cosmology and Astroparticle Physics}, 2020(01):024, 2020.

\bibitem{Carr2022}
Bernard Carr and Florian K{\"u}hnel.
\newblock Primordial black holes as dark matter candidates.
\newblock {\em SciPost Physics Lecture Notes}, page 048, 2022.

\bibitem{Villanueva2021}
Pablo Villanueva-Domingo, Olga Mena, and Sergio Palomares-Ruiz.
\newblock A brief review on primordial black holes as dark matter.
\newblock {\em Frontiers in Astronomy and Space Sciences}, 8:681084, 2021.

\bibitem{Garcia2017}
Juan Garc{\'\i}a-Bellido.
\newblock Massive primordial black holes as dark matter and their detection with gravitational waves.
\newblock In {\em Journal of Physics: Conference Series}, volume 840, page 012032. IOP Publishing, 2017.

\bibitem{nelson2021bouncing}
Nelson Pinto-Neto.
\newblock Bouncing quantum cosmology.
\newblock {\em Universe}, 7(4):110, 2021.

\bibitem{PatrickReview2}
Robert Brandenberger and Patrick Peter.
\newblock Bouncing cosmologies: Progress and problems.
\newblock {\em Foundations of Physics}, 47(6):797--850, feb 2017.

\bibitem{covariant_bardeen}
S.~D.~P. {Vitenti}, F.~T. {Falciano}, and N.~{Pinto-Neto}.
\newblock {Covariant Bardeen perturbation formalism}.
\newblock {\em Phys. Rev. D}, 89(10):103538, May 2014.

\bibitem{Gasperini1993}
M.~Gasperini and G.~Veneziano.
\newblock Inflation, deflation, and frame-independence in string cosmology.
\newblock {\em Mod. Phys. Lett. A}, 8:3701--3713, 1993.

\bibitem{Gasperini1994}
M.~Gasperini and G.~Veneziano.
\newblock Dilaton production in string cosmology.
\newblock {\em Phys. Rev. D}, 50:2519--2540, 8 1994.

\bibitem{Lyth}
David~H Lyth.
\newblock The primordial curvature perturbation in the ekpyrotic universe.
\newblock {\em Physics Letters B}, 524(1-2):1--4, 2002.

\bibitem{Finelli}
Fabio Finelli and Robert Brandenberger.
\newblock Generation of a scale-invariant spectrum of adiabatic fluctuations in cosmological models with a contracting phase.
\newblock {\em Physical Review D}, 65(10):103522, 2002.

\bibitem{Wands1999}
D.~Wands.
\newblock Duality invariance of cosmological perturbation spectra.
\newblock {\em Phys. Rev. D}, 60(2):023507, 7 1999.

\bibitem{Brandenberger2001}
R.~Brandenberger and F.~Finelli.
\newblock On the spectrum of fluctuations in an effective field theory of the ekpyrotic universe.
\newblock {\em J. High Energy Phys.}, 11:56, 11 2001.

\bibitem{Peter2002}
Patrick Peter and Nelson Pinto-Neto.
\newblock {Primordial perturbations in a nonsingular bouncing universe model}.
\newblock {\em Phys. Rev. D}, 66(6):063509, sep 2002.

\bibitem{Hwang2002}
J.~Hwang and H.~Noh.
\newblock Non-singular big-bounces and evolution of linear fluctuations.
\newblock {\em Phys. Rev. D}, 65:124010, 2002.

\bibitem{Vitenti2012}
Sandro Dias~Pinto Vitenti and Nelson Pinto-Neto.
\newblock Large adiabatic scalar perturbations in a regular bouncing universe.
\newblock {\em Physical Review D}, 85(2), January 2012.

\bibitem{Vitenti2013}
SDP Vitenti, FT~Falciano, and N~Pinto-Neto.
\newblock Quantum cosmological perturbations of generic fluids in quantum universes.
\newblock {\em Physical Review D}, 87(10):103503, 2013.

\bibitem{PatrickReview1}
Diana Battefeld and Patrick Peter.
\newblock A critical review of classical bouncing cosmologies.
\newblock {\em Physics Reports}, 571:1--66, apr 2015.

\bibitem{nelson_peter_bouncing_original}
Patrick Peter, Emanuel~JC Pinho, and Nelson Pinto-Neto.
\newblock Noninflationary model with scale invariant cosmological perturbations.
\newblock {\em Physical Review D}, 75(2):023516, 2007.

\bibitem{loop_quantum_gravity_perturbations_application}
Ivan Agullo, Javier Olmedo, and V~Sreenath.
\newblock Observational consequences of bianchi i spacetimes in loop quantum cosmology.
\newblock {\em Physical Review D}, 102(4):043523, 2020.

\bibitem{loop_phenomenology}
Mairi Sakellariadou.
\newblock Phenomenology of loop quantum cosmology.
\newblock In {\em Journal of Physics: Conference Series}, volume 222, page 012027. IOP Publishing, 2010.

\bibitem{Carr2011}
BJ~Carr and AA~Coley.
\newblock Persistence of black holes through a cosmological bounce.
\newblock {\em International Journal of Modern Physics D}, 20(14):2733--2738, 2011.

\bibitem{Corman2022}
Maxence Corman, William~E East, and Justin~L Ripley.
\newblock Evolution of black holes through a nonsingular cosmological bounce.
\newblock {\em Journal of Cosmology and Astroparticle Physics}, 2022(09):063, 2022.

\bibitem{Chen2017}
Jie-Wen Chen, Junyu Liu, Hao-Lan Xu, and Yi-Fu Cai.
\newblock Tracing primordial black holes in nonsingular bouncing cosmology.
\newblock {\em Physics Letters B}, 769:561--568, 2017.

\bibitem{Chen2023}
Jie-Wen Chen, Mian Zhu, Sheng-Feng Yan, Qing-Qing Wang, and Yi-Fu Cai.
\newblock Enhance primordial black hole abundance through the non-linear processes around bounce point.
\newblock {\em Journal of Cosmology and Astroparticle Physics}, 2023(01):015, 2023.

\bibitem{Quintin2016}
Jerome Quintin and Robert~H. Brandenberger.
\newblock Black hole formation in a contracting universe.
\newblock {\em Journal of Cosmology and Astroparticle Physics}, 2016(11):029–029, November 2016.

\bibitem{Banerjee2022}
Shreya Banerjee, Theodoros Papanikolaou, and Emmanuel~N Saridakis.
\newblock Constraining f (r) bouncing cosmologies through primordial black holes.
\newblock {\em Physical Review D}, 106(12):124012, 2022.

\bibitem{Papanikolaou2024}
Theodoros Papanikolaou, Shreya Banerjee, Yi-Fu Cai, Salvatore Capozziello, and Emmanuel~N Saridakis.
\newblock Primordial black holes and induced gravitational waves in non-singular matter bouncing cosmology.
\newblock {\em arXiv preprint arXiv:2404.03779}, 2024.

\bibitem{vitenti2012large}
Sandro Dias~Pinto Vitenti and Nelson Pinto-Neto.
\newblock Large adiabatic scalar perturbations in a regular bouncing universe.
\newblock {\em Physical Review D}, 85(2):023524, 2012.

\bibitem{nelson2000bohm}
Nelson Pinto-Neto.
\newblock Quantum cosmology: how to interpret and obtain results.
\newblock {\em Brazilian Journal of Physics}, 30:330--345, 2000.

\bibitem{Tolman1934}
Richard~C. Tolman.
\newblock {\em Relativity, Thermodynamics and Cosmology}.
\newblock Dover, 1934.

\bibitem{Lemaitre1933}
Georges Lema{\^\i}tre.
\newblock L'univers en expansion.
\newblock In {\em Annales de la Soci{\'e}t{\'e} scientifique de Bruxelles}, volume~53, page~51, 1933.

\bibitem{fluidgeral}
S.~D.~P. Vitenti, F.~T. Falciano, and N.~Pinto-Neto.
\newblock Quantum cosmological perturbations of generic fluids in quantum universes.
\newblock {\em Physical Review D}, 87(10), may 2013.

\bibitem{nelsonhamiltonian}
Nelson~Pinto Neto.
\newblock {\em Hamiltonian formulation of General Relativity and applications}.
\newblock Cadernos de Astrofísica, Cosmologia e Gravitação. PPGCosmo, 2020.

\bibitem{halliwell1990introductory}
Jonathan~J Halliwell.
\newblock Introductory lectures on quantum cosmology.
\newblock {\em Introductory lectures on quantum cosmology}, 1990.

\bibitem{nelson_bohm2023}
N.~{Pinto-Neto} and J.~C. {Fabris}.
\newblock {Quantum cosmology from the de Broglie-Bohm perspective}.
\newblock {\em Classical and Quantum Gravity}, 30(14):143001, July 2013.

\bibitem{dewitt1967}
Bryce~S. DeWitt.
\newblock Quantum theory of gravity. {I}. the canonical theory.
\newblock {\em Phys. Rev.}, 160:1113--1148, 1967.

\bibitem{patrick_time_review}
Claus Kiefer and Patrick Peter.
\newblock Time in quantum cosmology.
\newblock {\em Universe}, 8(1):36, 2022.

\bibitem{bianchi_time}
Przemys{\l}aw Ma{\l}kiewicz, Patrick Peter, and SDP Vitenti.
\newblock Quantum empty bianchi i spacetime with internal time.
\newblock {\em Physical Review D}, 101(4):046012, 2020.

\bibitem{mukhanov2005physical}
Viatcheslav Mukhanov.
\newblock {\em Physical foundations of cosmology}.
\newblock Cambridge university press, 2005.

\bibitem{Peter2016a}
P.~{Peter}, N.~{Pinto-Neto}, and S.~D.~P. {Vitenti}.
\newblock {Quantum cosmological perturbations of multiple fluids}.
\newblock {\em Phys. Rev. D}, 93(2):023520, 2016.

\bibitem{Bardeen1980}
J.~M. Bardeen.
\newblock Gauge-invariant cosmological perturbations.
\newblock {\em Phys. Rev. D}, 22:1882--1905, 10 1980.

\bibitem{Mukhanov1992}
V.~F. Mukhanov, H.~A. Feldman, and R.~H. Brandenberger.
\newblock Theory of cosmological perturbations.
\newblock {\em Phys. Rep.}, 215:203--333, 6 1992.

\bibitem{mukhanov1981quantum}
Viatcheslav~F Mukhanov and GV~Chibisov.
\newblock Quantum fluctuations and a nonsingular universe.
\newblock {\em ZhETF Pisma Redaktsiiu}, 33:549--553, 1981.

\bibitem{hawking1982quantum_fluctuations}
Stephen~W Hawking.
\newblock The development of irregularities in a single bubble inflationary universe.
\newblock {\em Physics Letters B}, 115(4):295--297, 1982.

\bibitem{Peter2005}
P.~Peter, E.~Pinho, and N.~Pinto-Neto.
\newblock Tensor perturbations in quantum cosmological backgrounds.
\newblock {\em J. Cosmol. Astropart. Phys.}, 7:14, 7 2005.

\bibitem{wald1994quantum}
Robert~M Wald.
\newblock {\em Quantum field theory in curved spacetime and black hole thermodynamics}.
\newblock University of Chicago press, 1994.

\bibitem{birrell1984quantum}
Nicholas~David Birrell and Paul Charles~William Davies.
\newblock {\em Quantum fields in curved space}.
\newblock Cambridge university press, 1984.

\bibitem{mukhanov2007introduction}
V.~Mukhanov and S.~Winitzki.
\newblock {\em Introduction to Quantum Effects in Gravity}.
\newblock Cambridge University Press, 2007.

\bibitem{vacuum2022}
Mariana Penna-Lima, Nelson Pinto-Neto, and Sandro~DP Vitenti.
\newblock New formalism to define vacuum states for scalar fields in curved space-times.
\newblock {\em arXiv preprint arXiv:2207.08270}, 2022.

\bibitem{Vitenti2014}
Sandro Dias~Pinto Vitenti and M.~{Penna-Lima}.
\newblock {NumCosmo: Numerical Cosmology}.
\newblock Astrophysics Source Code Library ascl:1408.013, aug 2014.

\bibitem{Baugh}
Carlton Baugh and P~Murdin.
\newblock Correlation function and power spectra in cosmology.
\newblock {\em Encyclopedia of Astronomy and Astrophysics,(IOP, London, UK, 2006)}, 2006.

\bibitem{planck_inflation_constraints}
Planck Collaboration.
\newblock {Planck 2018 results. X. Constraints on inflation}.
\newblock {\em A\&A}, 641:A10, September 2020.

\bibitem{Wang2014}
Yi~Wang and Wei Xue.
\newblock Inflation and alternatives with blue tensor spectra.
\newblock {\em Journal of Cosmology and Astroparticle Physics}, 2014(10):075–075, October 2014.

\bibitem{Cai2015}
Yi-Fu Cai, Jinn-Ouk Gong, Shi Pi, Emmanuel~N Saridakis, and Shang-Yu Wu.
\newblock On the possibility of blue tensor spectrum within single field inflation.
\newblock {\em Nuclear Physics B}, 900:517--532, 2015.

\bibitem{Kuroyanagi2021}
Sachiko Kuroyanagi, Tomo Takahashi, and Shuichiro Yokoyama.
\newblock Blue-tilted inflationary tensor spectrum and reheating in the light of nanograv results.
\newblock {\em Journal of Cosmology and Astroparticle Physics}, 2021(01):071, 2021.

\bibitem{Wu2023}
Yu-Mei Wu, Zu-Cheng Chen, and Qing-Guo Huang.
\newblock Search for stochastic gravitational-wave background from massive gravity in the nanograv 12.5-year dataset.
\newblock {\em Physical Review D}, 107(4):042003, 2023.

\bibitem{Q2C2}
Nelson Pinto-Neto, Grasiele Santos, and Ward Struyve.
\newblock Quantum-to-classical transition of primordial cosmological perturbations in de broglie--bohm quantum theory.
\newblock {\em Phys. Rev. D}, 85:083506, Apr 2012.

\bibitem{Musco2019}
Ilia Musco.
\newblock Threshold for primordial black holes: Dependence on the shape of the cosmological perturbations.
\newblock {\em Physical Review D}, 100(12), dec 2019.

\bibitem{Bardeen1986statistics}
James~M Bardeen, JR~Bond, Nick Kaiser, and AS~Szalay.
\newblock The statistics of peaks of gaussian random fields.
\newblock {\em Astrophysical Journal, Part 1 (ISSN 0004-637X), vol. 304, May 1, 1986, p. 15-61. SERC-supported research.}, 304:15--61, 1986.

\bibitem{Martin2014}
J.~{Martin}, C.~{Ringeval}, and V.~{Vennin}.
\newblock {Encyclop{\ae}dia Inflationaris}.
\newblock {\em Physics of the Dark Universe}, 5:75--235, December 2014.

\bibitem{Mathematica}
Wolfram~Research{,} Inc.
\newblock Mathematica, {V}ersion 14.0.
\newblock Champaign, IL, 2024.

\bibitem{Dey2023}
Dipanjan Dey, NT~Layden, AA~Coley, and Pankaj~S Joshi.
\newblock The equilibrium condition in gravitational collapse and its application to a cosmological scenario.
\newblock {\em arXiv preprint arXiv:2303.16789}, 2023.

\bibitem{Carr2021}
Bernard Carr, Kazunori Kohri, Yuuiti Sendouda, and Jun’ichi Yokoyama.
\newblock Constraints on primordial black holes.
\newblock {\em Reports on Progress in Physics}, 84(11):116902, 2021.

\bibitem{press1974}
William~H Press and Paul Schechter.
\newblock Formation of galaxies and clusters of galaxies by self-similar gravitational condensation.
\newblock {\em The Astrophysical Journal}, 187:425--438, 1974.

\bibitem{Cai2023}
Yi-Fu Cai, Chengfeng Tang, Geyu Mo, Sheng-Feng Yan, Chao Chen, Xiao-Han Ma, Bo~Wang, Wentao Luo, Damien Easson, and Antonino Marciano.
\newblock Primordial black hole mass functions as a probe of cosmic origin.
\newblock {\em arXiv preprint arXiv:2301.09403}, 2023.

\bibitem{Bi2024}
Xiao-Ming Bi, Lu~Chen, and Ke~Wang.
\newblock Primordial black hole mass function with mass gap.
\newblock {\em Monthly Notices of the Royal Astronomical Society}, 527(2):3962--3967, 2024.

\bibitem{Carr2016}
Bernard Carr, Florian Kühnel, and Marit Sandstad.
\newblock Primordial black holes as dark matter.
\newblock {\em Physical Review D}, 94(8), oct 2016.

\bibitem{Armendariz2014}
Cristian Armendariz-Picon and Jayanth~T Neelakanta.
\newblock How cold is cold dark matter?
\newblock {\em Journal of Cosmology and Astroparticle Physics}, 2014(03):049, 2014.

\bibitem{Gonccalves2000}
S{\'e}rgio~MCV Gon{\c{c}}alves.
\newblock Black hole formation from massive scalar field collapse in the einstein--de sitter universe.
\newblock {\em Physical Review D}, 62(12):124006, 2000.

\bibitem{Vitenti2014covariant}
SDP Vitenti, FT~Falciano, and N~Pinto-Neto.
\newblock Covariant bardeen perturbation formalism.
\newblock {\em Physical Review D}, 89(10):103538, 2014.

\end{thebibliography}

\end{document}